\documentclass{article}
\usepackage{spconf}
\usepackage{amsmath,graphicx}
\usepackage{amsfonts} 

\usepackage{enumitem}
\setlist{nosep, leftmargin=14pt}

\usepackage{amsmath, amsthm}
\usepackage{amssymb}
\usepackage{graphicx}
\usepackage{xcolor}
\usepackage{footnote}
\usepackage{soul}

\usepackage{mathtools} 
\usepackage[colorlinks=true, allcolors=blue, bookmarks=false]{hyperref}
\usepackage{url}

\newcommand{\SO}{\mathsf{SO}}
\newcommand{\R}{\mathbb{R}}
\renewcommand{\O}{\mathsf{O}}
\newcommand{\C}{\mathbb{C}}

\newcommand\blfootnote[1]{%
  \begingroup
  \renewcommand\thefootnote{}\footnote{#1}%
  \addtocounter{footnote}{-1}%
  \endgroup
}

\usepackage{color}

\numberwithin{equation}{section}

\title{$\SO(3)$-invariant PCA with application to molecular data}

\name{\parbox{\textwidth-5mm}{\centering
Michael Fraiman$^{\star}$ \qquad Paulina Hoyos$^{\dagger}$
\qquad Tamir Bendory$^{\P}$
\qquad 
Joe Kileel$^{\dagger}$ \\  Oscar Mickelin$^{\ddagger}$
 \qquad Nir Sharon$^{\star}$ 
 \qquad Amit Singer$^{\S}$
}}
 \address{$^{\star}$ School of Mathematical Sciences, Tel Aviv University, Israel \\ 
    $^{\dagger}$  Department of Mathematics, UT Austin, USA \\ 
    $^{\P}$ School of Electrical and Computer Engineering, Tel Aviv University, Israel \\ 
    $^{\ddagger}$  Yau Mathematical Sciences Center, Tsinghua University, China\\ 
    $^{\S}$ Program in Applied and Computational Mathematics\\and Department of Mathematics, Princeton University, USA 
    }
\begin{document}
\maketitle
\begin{abstract}
Principal component analysis (PCA) is a fundamental technique for dimensionality reduction and denoising; however, its application to three-dimensional data with arbitrary orientations---common in structural biology---presents significant challenges. A naive approach requires augmenting the dataset with many rotated copies of each sample, incurring prohibitive computational costs. In this paper, we extend PCA to 3D volumetric datasets with unknown orientations by developing an efficient and principled framework for $\SO(3)$-invariant PCA that implicitly accounts for all rotations without explicit data
augmentation. By exploiting underlying algebraic structure, we demonstrate that the computation involves only the square root of the total number of covariance entries, resulting in a substantial reduction in complexity. We validate the method on real-world molecular datasets, demonstrating its effectiveness and opening up new possibilities for large-scale, high-dimensional reconstruction problems.
\end{abstract}

\section{Introduction}
\label{sec:intro}

\blfootnote{M.F. and P.H. contributed equally to this work.}
Principal component analysis (PCA) is a fundamental technique in data science and statistics, especially when dealing with high-dimensional datasets. 
By extracting the leading eigenvectors of the covariance matrix, PCA identifies directions of maximum variability and projects the data onto a lower-dimensional space. This simple and interpretable method is widely used for denoising, dimensionality reduction, visualization, feature extraction, and compression, while often preserving the essential structure of the data~\cite{jolliffe2016principal,jolliffe2011principal}.

In some scenarios, the orientation of the data samples is arbitrary or random. A notable example, and the focus of this paper, arises when the samples are three-dimensional functions, each observed up to an unknown rotation.  
As in~\cite{zhao2016fast, rosen2024ginvariant},
rather than using the costly naive fix---augmenting each sample with many rotated copies to handle random orientations---we exploit the underlying algebraic structure and group symmetries.  This results in a numerically accurate method that implicitly accounts for all rotations, while greatly reducing computation.
In particular, we prove that, under a suitable representation, the $\mathrm{SO}(3)$-invariant covariance matrix is low-dimensional and its eigenvectors can be computed efficiently. The problem is formally stated in Section~\ref{sec:problem_formulation}, and the derivation of the $\SO(3)$-invariant PCA is presented in Section~\ref{sec:pca}.

The primary motivation of this paper arises from molecular datasets in  structural biology applications,
    including single-particle cryo-electron microscopy and cryo-electron tomography~\cite{nogales2016development,cheng2018single,watson2024advances}.
These problems are computationally demanding, involving the reconstruction of 3D structures from massive, high-dimensional, noisy datasets~\cite{bendory2020single}.
Although PCA has been highly effective in various stages of cryo-EM image analysis, its use has largely been limited to 2D raw images, leaving a gap in methodologies for  processing 3D volumes~\cite{zhao2016fast,bhamre2016denoising,van1981use}.
We extend PCA to 3D molecular data, enabling a principled approach for dimensionality reduction and denoising in volumetric reconstructions.
Section~\ref{sec:numerics} demonstrates the effectiveness of our method on available molecular datasets, and Section~\ref{sec:applications} outlines potential future applications.

\section{Problem Formulation}
\label{sec:problem_formulation}

Let $\phi \colon \R^3 \to \C$ be a function supported on the unit ball (e.g., the electro-static potential of a molecule, or its Fourier transform).  
We assume the function is well-approximated by the following finite-term expansion in spherical coordinates:
	\begin{equation} \label{eq:volume_expansion_app} 
        \phi(r, \theta, \varphi) =  \sum_{\ell = 0}^{L}\sum_{m=-\ell}^{\ell}\sum_{s=1}^{S(\ell)}f_{\ell ms} j_{\ell s}(r) Y_{\ell}^m(\theta,\varphi),
	\end{equation}
where $f_{\ell m s}\in\mathbb{C}$ are expansion coefficients. 
The basis functions $j_{\ell s}(r)Y_\ell^m(\theta, \varphi)$ are called \emph{ball harmonics} \cite{kileel2025fast}. 
Here $j_{\ell s}$ is the normalized spherical Bessel functions given by
$j_{\ell  s}(r) := \frac{4}{|j_{\ell+1}(u_{\ell  s})|}j_{\ell}(u_{\ell s} r)$, where
$j_{\ell}$ is the $\ell$-th spherical Bessel function of the first kind, and $u_{\ell s}$ is the $s$-th positive zero of $j_{\ell}$. 
The complex spherical harmonics $Y_{\ell}^m$ are given by  
$Y_{\ell}^m(\theta,\varphi) \coloneq \sqrt{\frac{2\ell+1}{4\pi}\cdot\frac{(\ell-m)!}{(\ell+m)!}}P_{\ell}^m(\cos\theta)e^{\iota m\varphi},$
where $P_{\ell}^m$ are the associated Legendre polynomials with the Condon--Shortley phase. 
Here ${S(\ell)}$ denotes the number of radial components retained at the angular degree $\ell$, which is determined by the Nyquist sampling rate of the discretized volume and decreases with $\ell$.
 Under Nyquist sampling on an $N \times N \times N$ grid, we have $S = \Theta (N)$ and $L = \mathcal{O}(N)$. 
Any square-integrable function on the unit ball can be represented in the form \eqref{eq:volume_expansion_app} as $L\rightarrow \infty$, and the finite expansion is common and useful to represent smooth, 3D volumes. See~\cite{kileel2025fast} for a fast algorithm to compute the expansion numerically.

A 3D rotation $R \in \SO(3)$ acts on the function $\phi \colon \R^3 \to \C$ via $(R \cdot \phi) (x) \coloneq \phi (R^T x)$ for $x \in \mathbb{R}^3$.
In spherical coordinates, the action is given by
\vspace*{-1mm}
\begin{align*}
    &(R\cdot \phi)(r, \theta, \varphi)\\
    &=  \sum_{\ell = 0}^{L}\sum_{m=-\ell}^{\ell}\sum_{s=1}^{S(\ell)}\left( \sum_{k=  -\ell}^{\ell}D^\ell_{mk}(R) f_{\ell  k s} \right) j_{\ell s}(r) Y_{\ell}^m(\theta,\varphi),
\end{align*}
where $D_{m k}^{\ell}(R)$ are the Wigner D-matrices.
This amounts to transforming the expansion coefficients of $\phi$  via
\vspace*{-1mm}
\begin{equation}\label{eq: expansion coefficients of rotated volumes}
R \cdot f_{\ell m s} = \sum_{k =-\ell}^{\ell} D_{m k}^{\ell}(R) f_{\ell k s}.
\end{equation}

We assume access to $n$ volumes of the form~\eqref{eq:volume_expansion_app}, each subjected to an arbitrary 3D rotation as described above and potentially corrupted by noise. Each volume resides in a high-dimensional space $\mathbb{R}^{D}$, 
where $D = \sum_{\ell=0}^{L} (2\ell+1)S(\ell)$.
The goal of PCA is to identify a lower-dimensional subspace isomorphic to $\mathbb{R}^d$ with $d \ll D$, such that the volumes can be well-approximated by their projections onto this subspace.
To this end, we compute the $\SO(3)$-invariant covariance matrix and extract its leading eigenvectors, as explained below.

\vspace*{-2mm}
\section{\texorpdfstring{$\SO(3)$}{SO(3)}-invariant Principal Components} \label{sec:pca}
\vspace*{-2mm}

Suppose we collect $n$ volumes $\{\phi^{(i)} \}_{i=1}^n$ of the form~\eqref{eq:volume_expansion_app}, and consider all of their possible 3D rotations. 
The sample mean of this $\SO(3)$-augmented collection is given by

\vspace*{-4mm}
\begin{align*}
    \phi_{\mathrm{mean}}(r, \theta, \varphi) &= \frac{1}{n} \sum_{i=1}^n \int_{\SO(3)} (R\cdot \phi^{(i)})(r, \theta, \varphi) dR \\
        & = \sum_{s=1}^{S(\ell)} \left( \frac{1}{n} \sum_{i=1}^n f^{(i)}_{00s}\right) j_{0 s}(r)Y_{0}^0(\theta,\varphi),
\end{align*}
where the last equality follows from Equation~\eqref{eq: expansion coefficients of rotated volumes} and  orthogonality relations of the Wigner D-matrices.
Since $Y_{0}^0(\theta, \varphi)$ is constant, $\phi_{\mathrm{mean}}$ is a radially symmetric function.

The full $\SO(3)$-invariant covariance is defined as
\begin{align*}
    \mathcal{C} = \frac{1}{n} \sum_{i=1}^n \,
    \smashoperator[r]{\int_{\SO(3)}}
    (R\cdot \phi^{(i)} - \phi_{\mathrm{mean}}) \overline{( R\cdot \phi^{(i)} - \phi_{\mathrm{mean}})} dR.
\end{align*}
To express $\mathcal{C}$ in the ball harmonics basis 
using the expansion coefficients $f^{(i)}_{\ell m s}$, we first note that centering the volumes amounts to updating the zero-frequency coefficients $f^{(i)}_{00 s}$
    to $f^{(i)}_{00 s} - \frac{1}{n} \sum_{i=1}^n f^{(i)}_{00s}$. 
    We then compute
\begin{align} \label{eq: SO(3)-invariant covariance coefficients}
    &\mathcal{C}_{(\ell, m, s), (\ell', m', s')} = \frac{1}{n}\sum_{i=1}^n \, \smashoperator[r]{\int_{\SO(3)}} (R \cdot f^{(i)}_{\ell m s}) \overline{ (R \cdot f^{(i)}_{\ell m' s'}) } dR \nonumber \\
    & =
      \frac{1}{n}\sum_{i=1}^n\sum_{k=-\ell}^{\ell}
      \smashoperator[r]{\sum_{k'=-\ell'}^{\ell'}}
        f^{(i)}_{\ell k s}\, \overline{f^{(i)}_{\ell' k' s'}}
      \smashoperator[lr]{\int\limits_{\SO(3)}}
        D^{\ell}_{m k}(R)\, \overline{D^{\ell'}_{m' k'}(R)}\, dR \nonumber \\
    &=
        \delta_{m m'} \cdot \delta_{\ell \ell'} \cdot \frac{1}{n} \cdot \frac{1}{2 \ell + 1}  \sum_{i=1}^n  \sum_{k=-\ell}^{\ell}f^{(i)}_{\ell k s}  \overline{f^{(i)}_{\ell' k s'}},    
\end{align}
where the last equality follows from  orthogonality relations of the  Wigner D-matrices.
Thus, 
if we set $C_{\ell} \in \C^{S(\ell) \times S(\ell)}$ as 
\begin{equation} \label{eq:covariancesmall}
    C_\ell (s,s') \coloneq \frac{1}{n} \cdot \frac{1}{2 \ell + 1}  \sum_{i=1}^n  \sum_{m=-\ell}^{\ell} f^{(i)}_{\ell m s}  \overline{f^{(i)}_{\ell m s'}},
\end{equation}
then 
\begin{equation}\label{eq:nice-one}
\mathcal{C} = \bigoplus_{\ell=0}^L \bigl( I_{2\ell+1} \otimes C_\ell \bigr).
\end{equation}
In other words, the matrix representation of $\mathcal{C}$ in \eqref{eq: SO(3)-invariant covariance coefficients} with respect to the ball harmonics basis is block-diagonal, with each block $C_\ell \in \mathbb{C}^{S(\ell) \times S(\ell)}$ occurring $(2 \ell + 1)$ times.
This reduction is possible due to the averaging over $\SO(3)$.
We note that formulas \eqref{eq:covariancesmall} and \eqref{eq:nice-one} extend results  of~\cite{zhao2016fast} from 2D to 3D.

Let $\{(v_{\ell s}, \lambda_{\ell s})\}_{s = 1}^{S(\ell)}$ be the eigenvector/eigenvalue pairs of the matrix $C_\ell$ in \eqref{eq:covariancesmall}.
Each $v_{\ell s}$ gives rise to a set $V_{\ell s} \coloneq \{ u_{\ell  s  m} \}_{m=-\ell}^{\ell}$, where $ u_{\ell s m} \coloneq e_m \otimes v_{\ell s}$ and $e_m$ is the $m$-th standard basis vector in  $\mathbb{R}^{2\ell + 1}$. 
    Each element of $V_{\ell s}$  is an eigenvector of $I_{2 \ell + 1}  \otimes C_\ell$ with the eigenvalue $\lambda_{\ell,s}$.
    By \eqref{eq:nice-one}, all such vectors, padded with zeros, are the eigenvectors of $\mathcal{C}$, and they constitute a basis of $\mathbb{R}^D$.
This lets us obtain the eigenvectors of $\mathcal{C}$ from those of $\{C_{\ell}\}_{\ell=0}^L$ at no extra cost.

We order the eigenvectors $\{u_{\ell_j s_j m_j} \}_{j=1}^D$ as follows.
    First, we order the sets $\{V_{\ell_{\mathbf{j}} s_{\mathbf{j}}}\}_{\mathbf{j}=1}^{D^\prime}$, where $D^\prime = \sum_{\ell=0}^L S(\ell)$, so the corresponding eigenvalues decrease.  
    Then, inside each $V_{\ell_{\mathbf{j}} s_{\mathbf{j}}}$ we order the vectors by~$m$.
The eigenvolumes
    $\mathbf{V} = \{\psi_j\}_{j=1}^D$
    are given by
    $\psi_{j} (r, \theta, \varphi) = \sum_{t=1}^{S(\ell_j)} (v_{\ell_j s_j})_t \, j_{\ell_j t} (r) \, Y_{\ell_j}^{m_j} (\theta, \varphi),$
        where $(\cdot)_t$ is the $t$-th entry of a vector.
Each volume $\phi$  decomposes as
${\phi = \sum_{j=1}^D \alpha_j \psi_j}$ with $\alpha_j := \sum_{t=1}^{S(\ell)} f_{\ell_j m_j t} \, \overline{(v_{\ell_j s_j })}_t$.  
Finally, the $\SO(3)$-invariant PCA 
approximation of $\phi$ is $ \sum_{j=1}^d \alpha_j \psi_j$ for $d \ll D$, with $d$ chosen by a rank criterion.

\textbf{Computational Complexity.}
Suppose each volume in the dataset $\{\phi^{(i)}\}_{i=1}^n$ is discretized on a cubic grid of side length $N$, so that each volume contains $N^3$ voxels. 
Recall $S := \max_\ell S(\ell)$.
Computation of the 
ball harmonics expansion coefficients takes ${\mathcal{O}(N^3  \log^2 (N))}$ operations~\cite{kileel2025fast}.
Computing the $\ell$-th block $C_\ell$ of the covariance matrix requires $\mathcal{O}(S^2 \ell)$ operations; summing over $\ell$,
    there are $\mathcal{O}(S^2 L^2)$  operations. 
The eigendecomposition takes $\mathcal{O}(S^3)$ operations for each block.
    Overall, the complexity of computing the (full) $\SO(3)$-invariant PCA basis is $\mathcal{O}(S^2 L^2 + S^3L)$. 
For comparison, a computation of the usual dense covariance of $\{\phi^{(i)}\}_{i=1}^n$ in the ball harmonics basis would require $\mathcal{O}(S^2 L^4)$ operations, and its eigendecomposition would cost $\mathcal{O}(S^3 L^6)$.
Under Nyquist sampling, our algorithm requires $\mathcal{O}(N^4)$ operations in total compared to $\mathcal{O}(N^9)$ for the usual computation.
Thus, our formulation of the $\SO(3)$-invariant covariance simultaneously achieves two objectives:
    it implicitly incorporates all rotated versions of the data, and it reduces the computational complexity of PCA by a factor of  about $N^5$.

 \textbf{$\O(3)$-Invariant PCA.} To construct an $\O(3)$-invariant covariance, we could augment the dataset with reflections of the volumes in the $xy$-plane, and then perform the $\SO(3)$-invariant procedure.  
This reflection transforms the coefficients by replacing $f_{\ell ms}$ with $(-1)^{\ell + m}f_{\ell ms}$. 
It turns out that the $\O(3)$-invariant covariance matrix is identical to \eqref{eq:nice-one}. 

\vspace*{-1mm}
\section{Numerical Results} \label{sec:numerics}

We test our framework on a dataset of 1,419 volumes from the Protein Data Bank~\cite{pdb} like in 
\cite{zhang2024moment}.  
Each volume is expanded in 
ball harmonics,  
by the implementation of~\cite{kileel2025fast} with a bandlimit $L = 20$ and resolutions $N = 64$, $128$,  and $256$ (i.e., each volume consists of $N^3$ voxels).  
Our code is at \path{https://github.com/MichaelFraiman/PCA\_SO-3}.

\begin{figure*}[p]
    \centering

    \begin{minipage}[b]{0.16\linewidth}
        \centering
        \includegraphics[width=\linewidth]{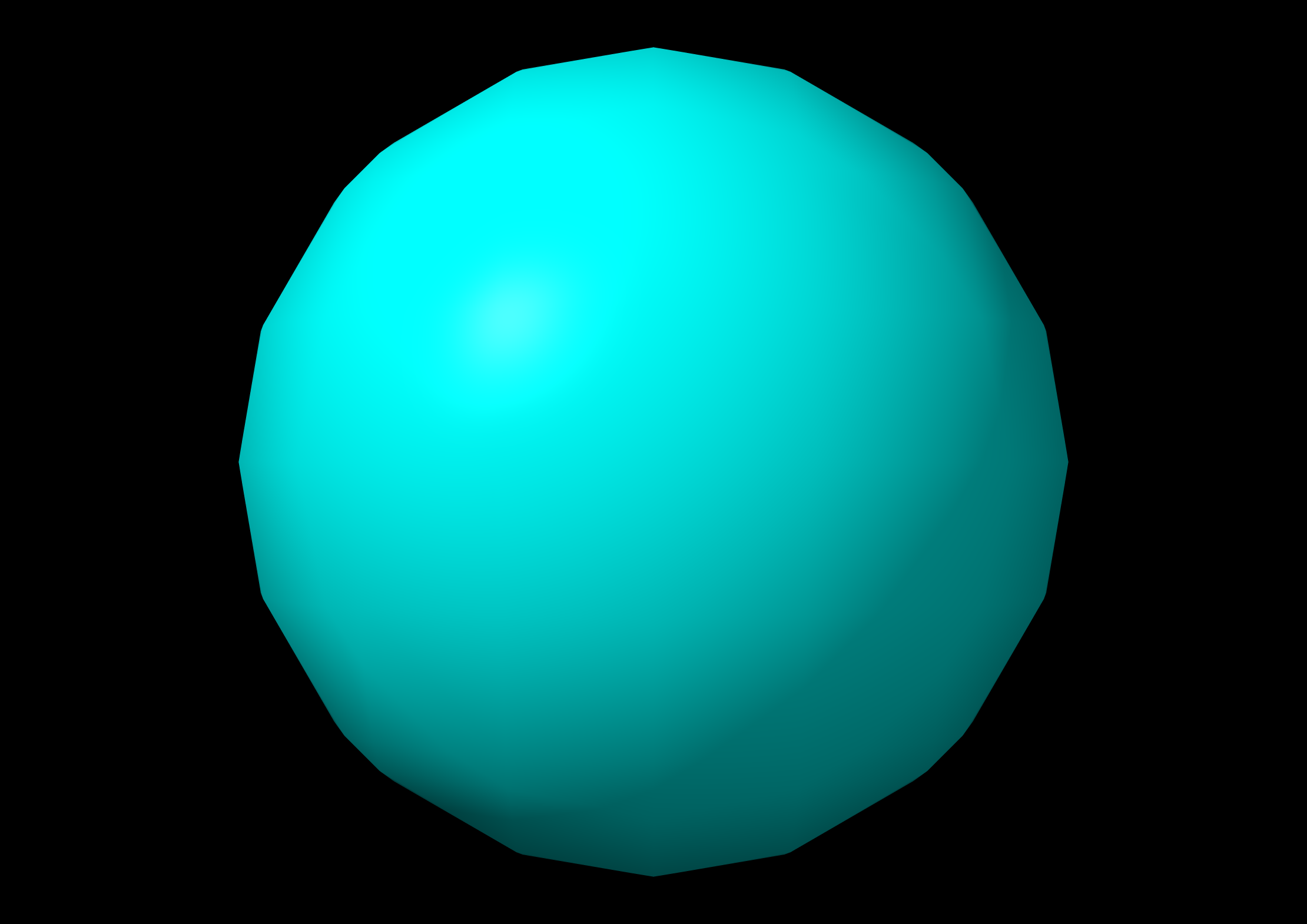}
        (a) $j=1$, $\mathbf{j} = 1$ ($s = 0$, $\ell = 0$, $m = 0$)
    \end{minipage}
    \hfill
    \begin{minipage}[b]{0.16\linewidth}
        \centering
        \includegraphics[width=\linewidth]{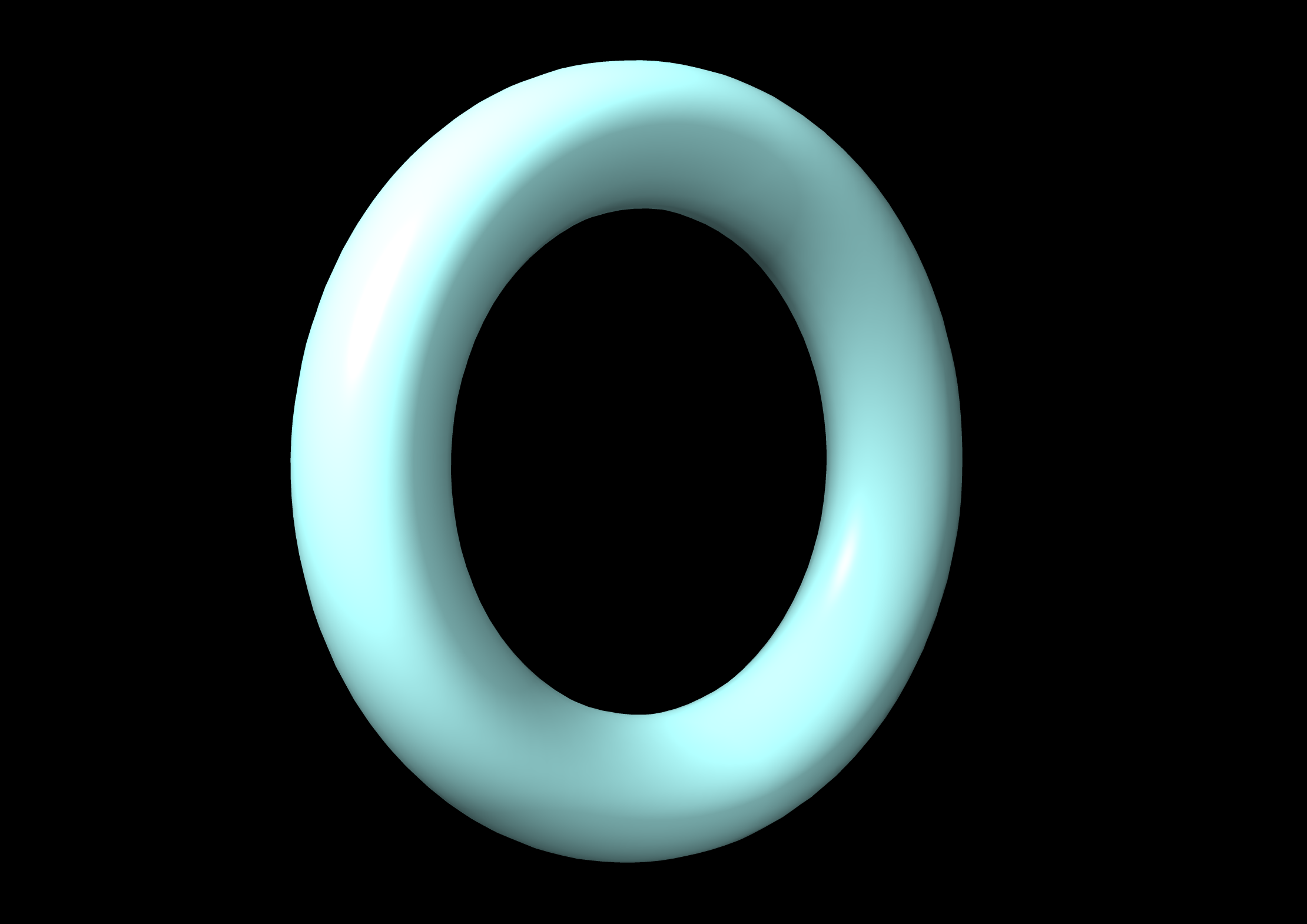}
        (b) $j=2$, $\mathbf{j} = 2$ ($s = 0$, $\ell = 2$, $m = -2$)
    \end{minipage}
    \hfill
    \begin{minipage}[b]{0.16\linewidth}
        \centering
        \includegraphics[width=\linewidth]{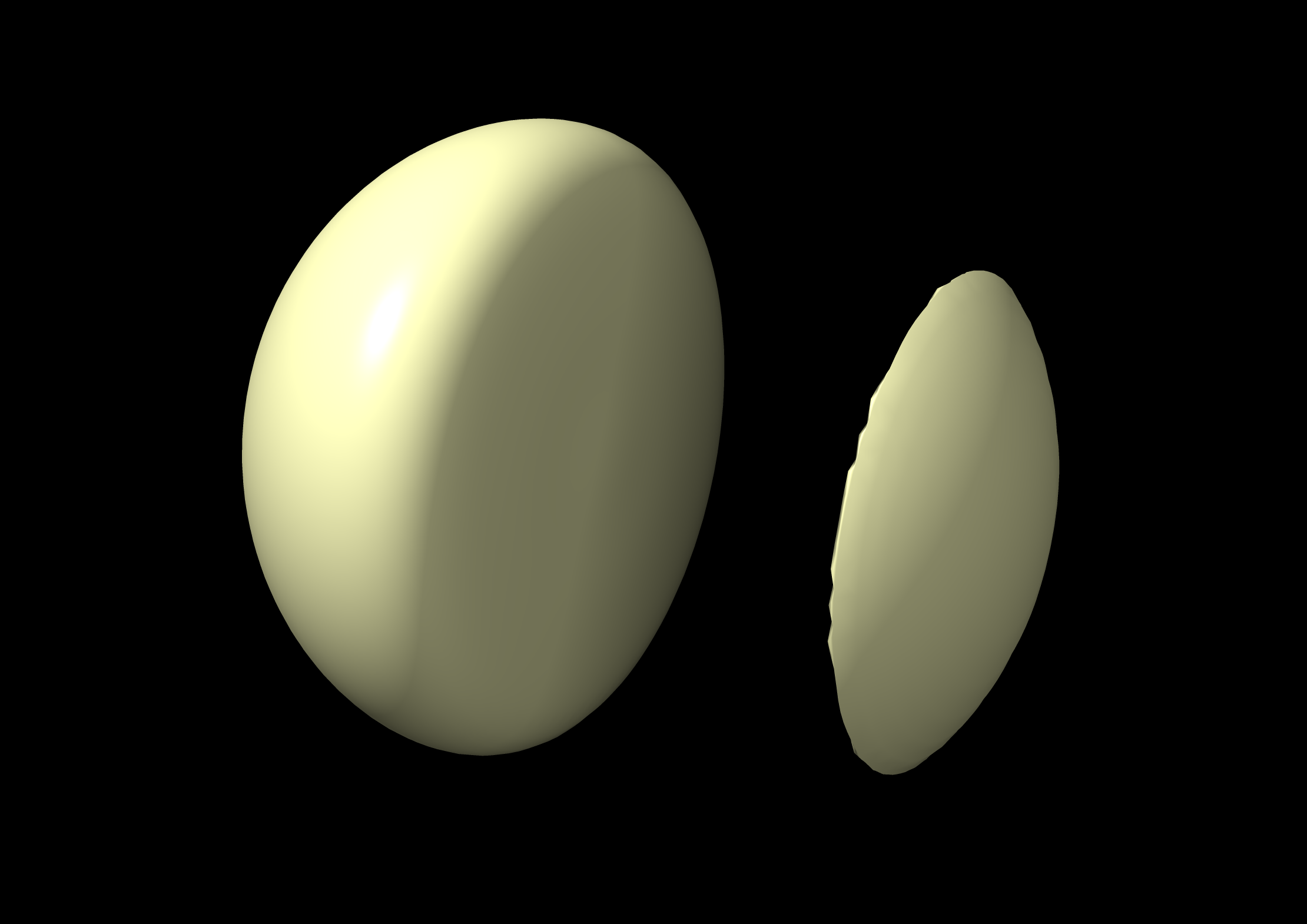}
        (e) $j=7$, $\mathbf{j} = 4$ ($s = 0$, $\ell = 1$, $m = -1$)
    \end{minipage}
    \hfill
    \begin{minipage}[b]{0.16\linewidth}
        \centering
        \includegraphics[width=\linewidth]{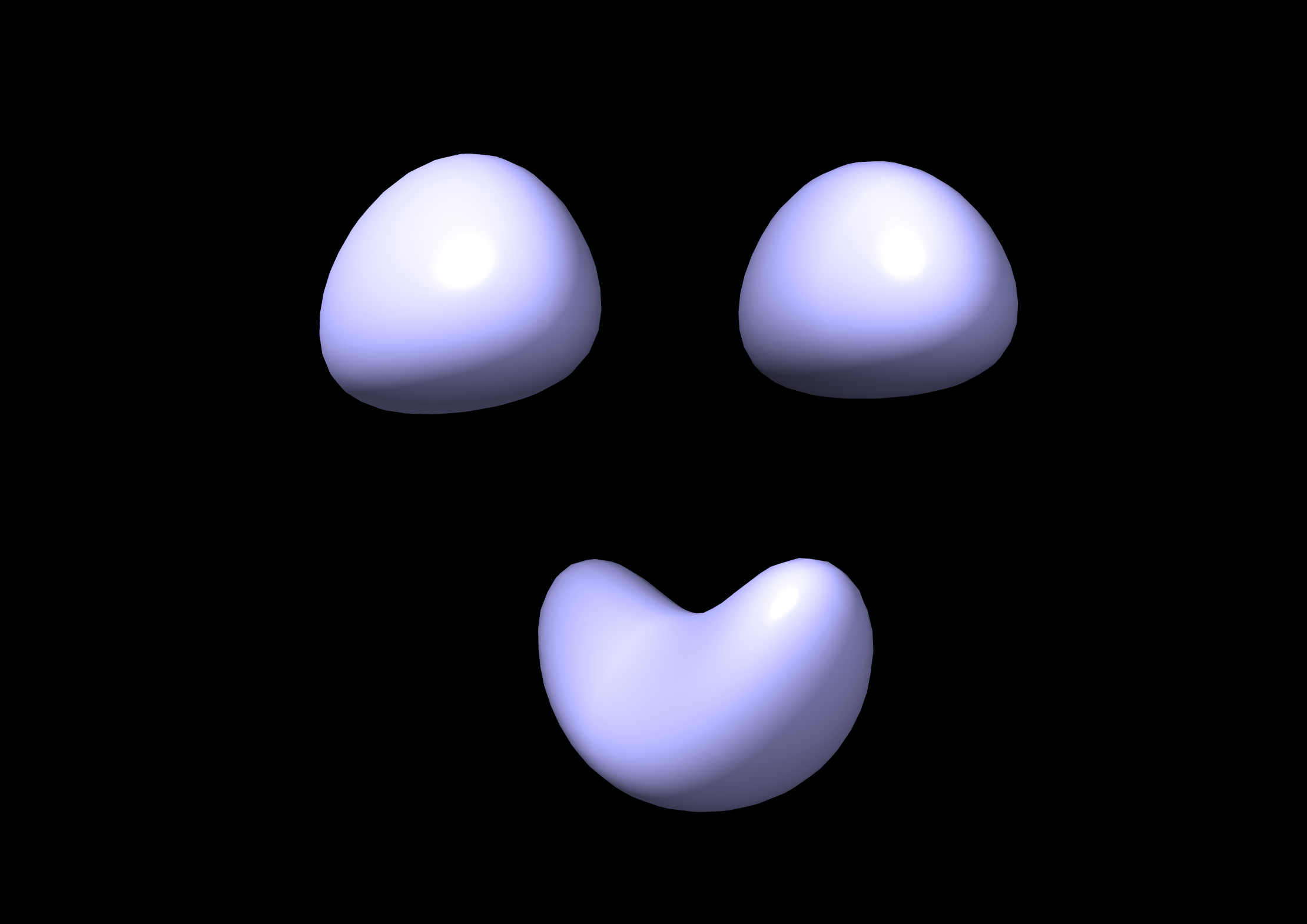}
        (c) $j=35$, $\mathbf{j} = 9$ ($s = 1$, $\ell = 1$, $m = 1$)
    \end{minipage}
    \hfill
    \begin{minipage}[b]{0.16\linewidth}
        \centering
        \includegraphics[width=\linewidth]{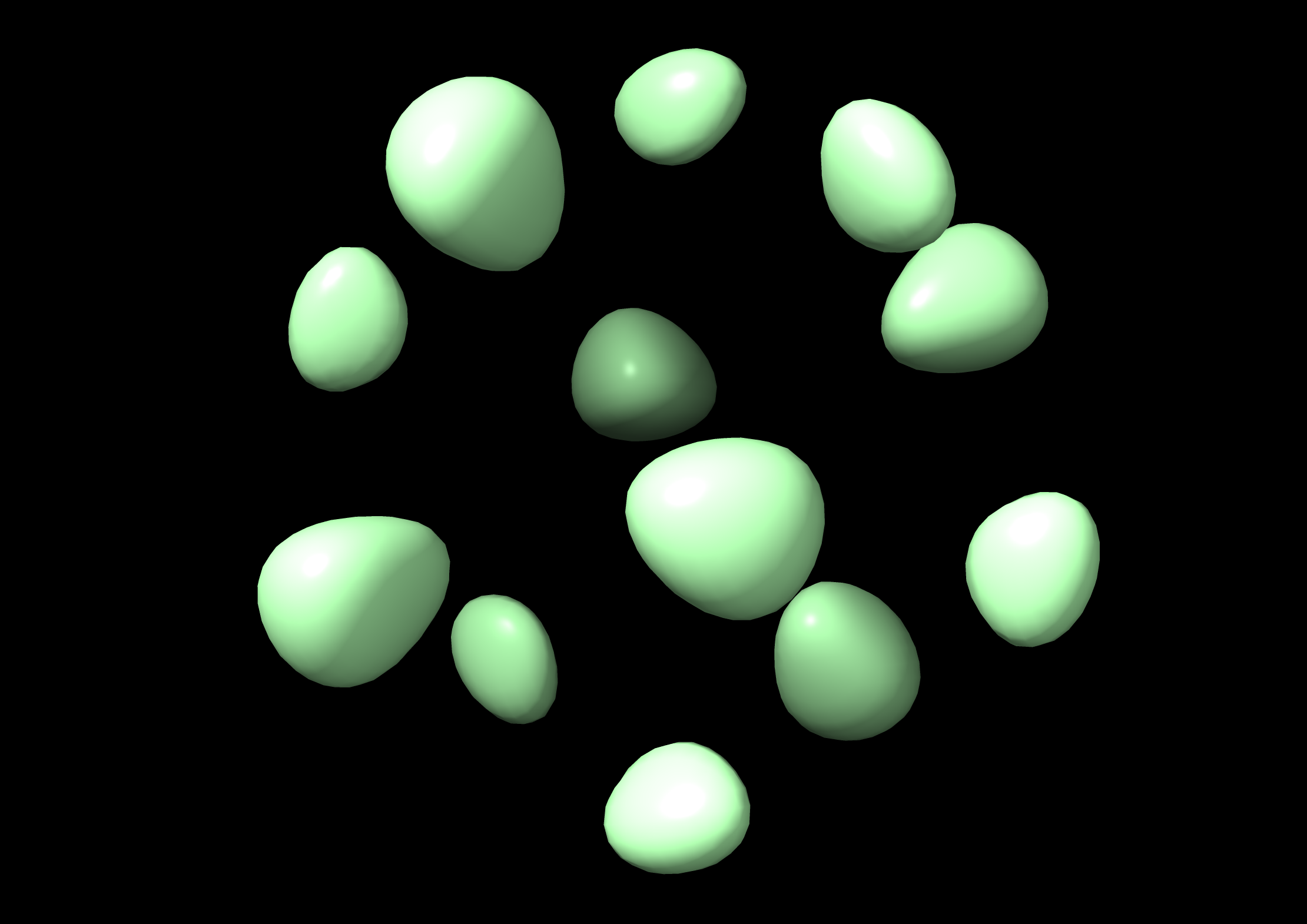}
        (d) $j=75$, $\mathbf{j} = 15$ ($s = 0$, $\ell = 6$, $m = 0$)
    \end{minipage}
    \hfill
    \begin{minipage}[b]{0.16\linewidth}
        \centering
        \includegraphics[width=\linewidth]{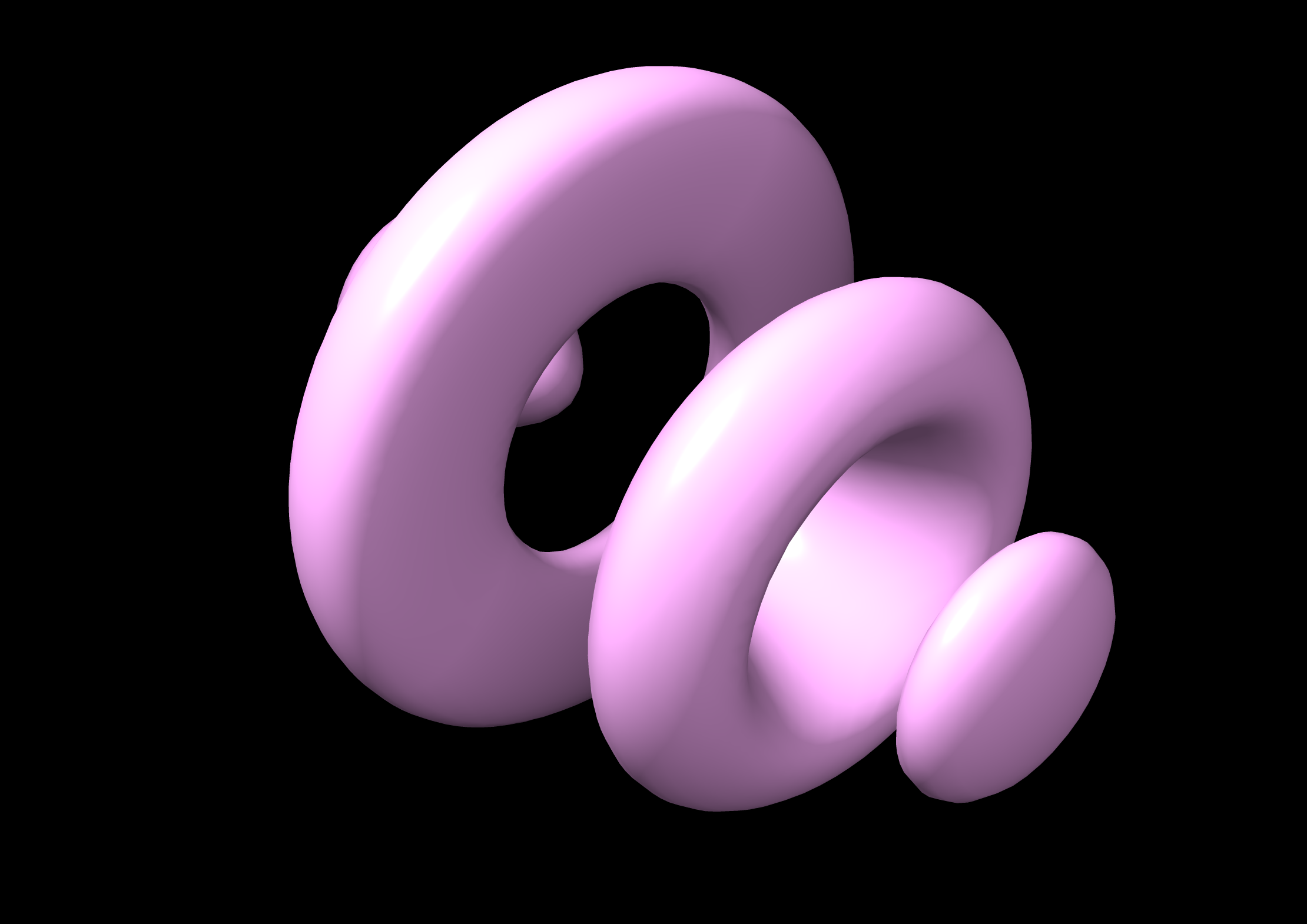}
        (f) $j=96$, $\mathbf{j} = 19$ ($s = 1$, $\ell = 5$, $m = -5$)
    \end{minipage}

    \caption{Eigenvolumes for $N = 64$.
    The notation is retained from  Section \ref{sec:pca}:
        $j$ stands for the rank of the eigenvolume $\psi_j$ in the ordering of ${u_{\ell_j s_j m_j}}$,
        while $\mathbf{j}$ stands for the rank of the eigenvalue $\lambda_{\ell_\mathbf{j} s_\mathbf{j}}$.
    }
    \label{fig:pcaeigenv}
\end{figure*}

\begin{figure*}[p]
    \centering
    \newlength{\lengthproteinspic}
    \setlength{\lengthproteinspic}{0.15\linewidth}
    \begin{minipage}[b]{\lengthproteinspic}
        \centering
        \includegraphics[width=\linewidth]{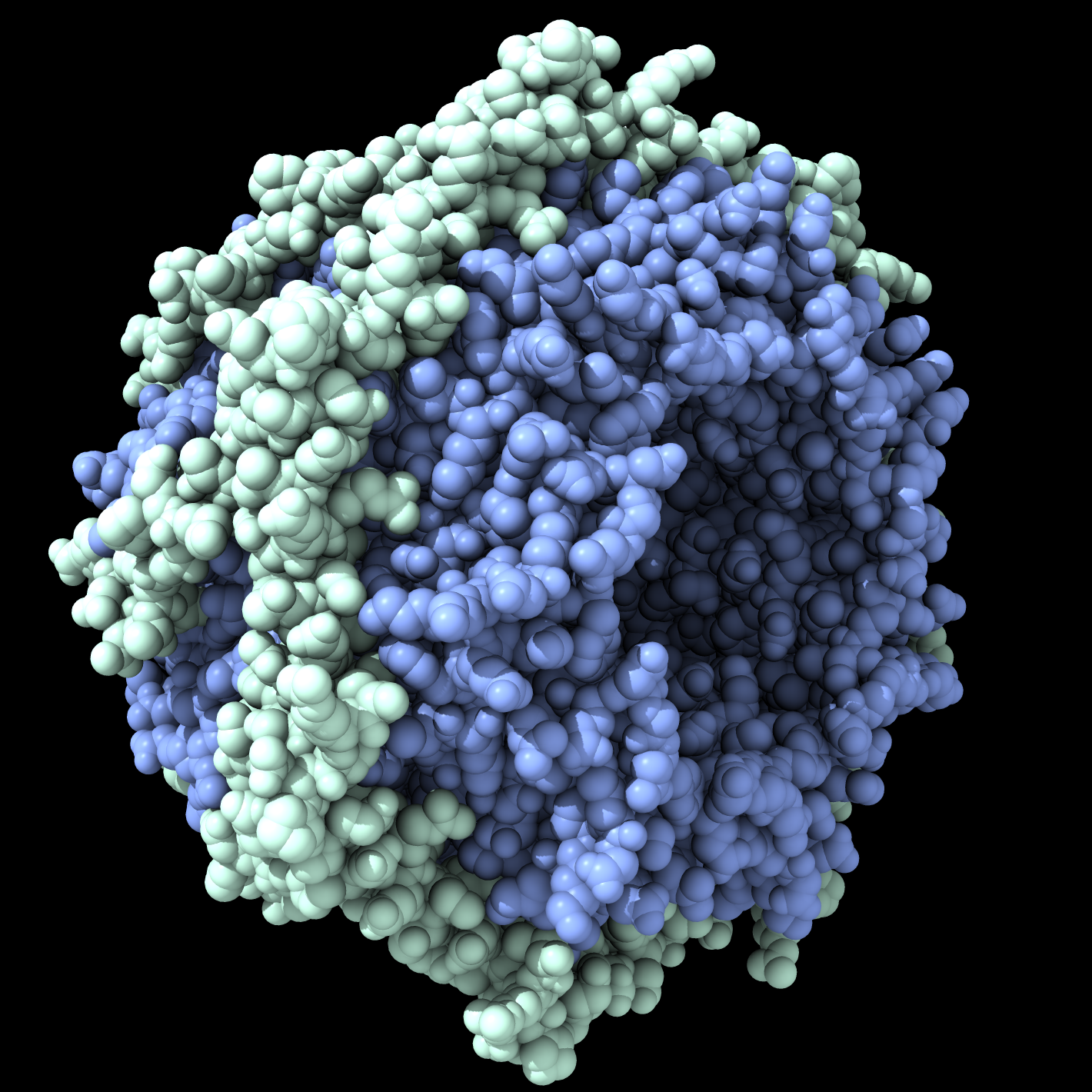}
        (a)
        \vspace{1mm}
        
        \includegraphics[width=\linewidth]{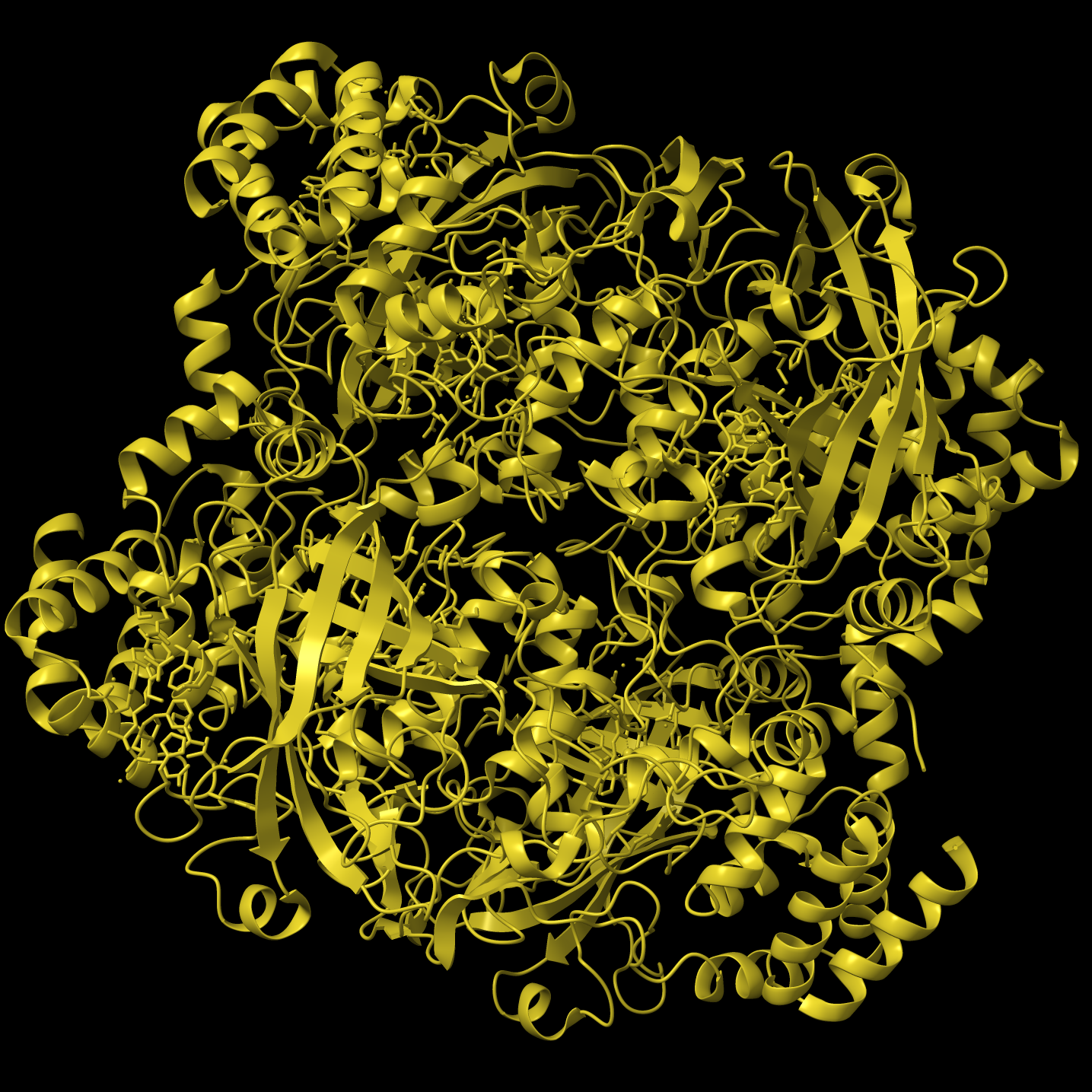}
    \end{minipage}
    \hfill
    \begin{minipage}[b]{\lengthproteinspic}
        \centering
        \includegraphics[width=\linewidth]{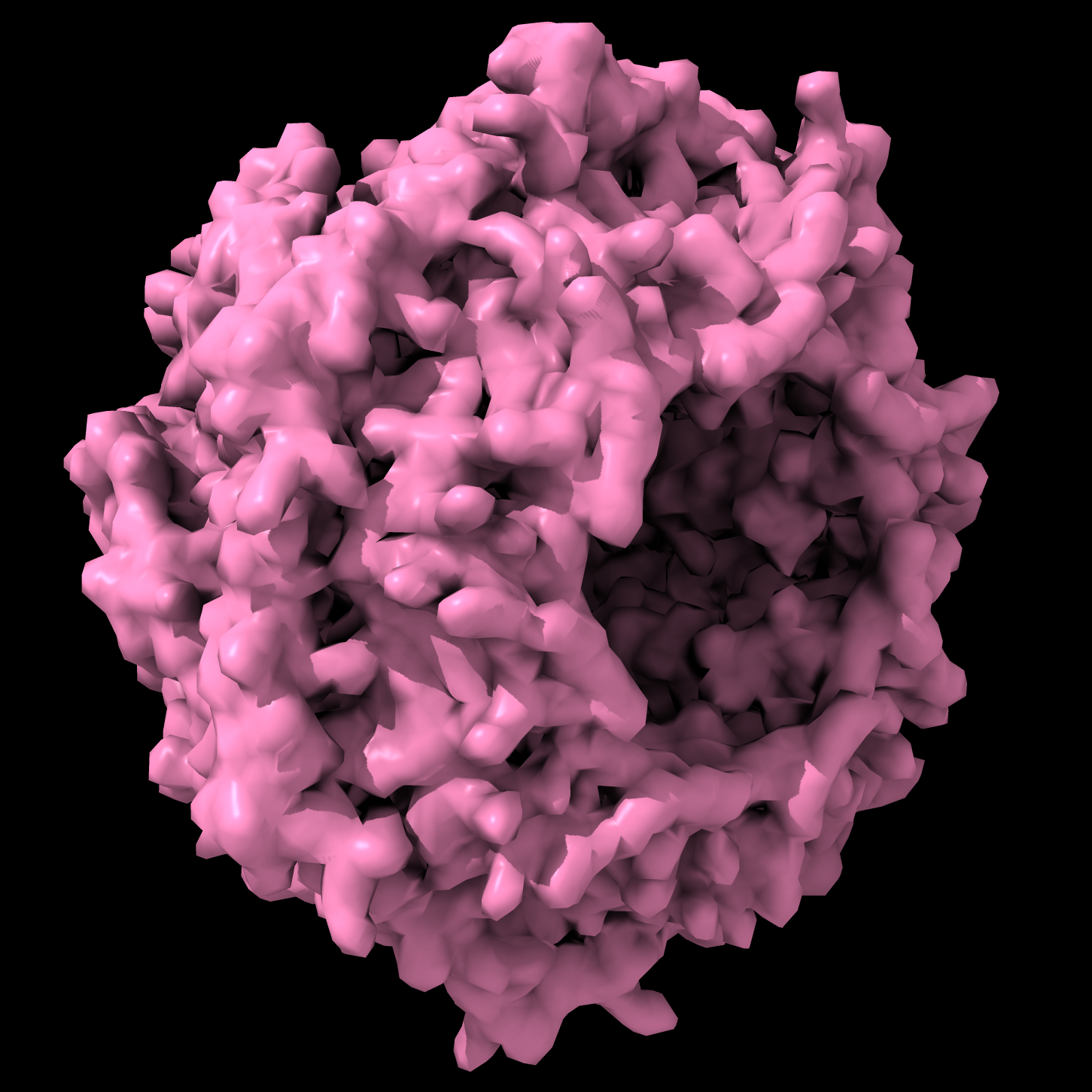}
        (b)
        \vspace{1mm}
        
        \includegraphics[width=\linewidth]{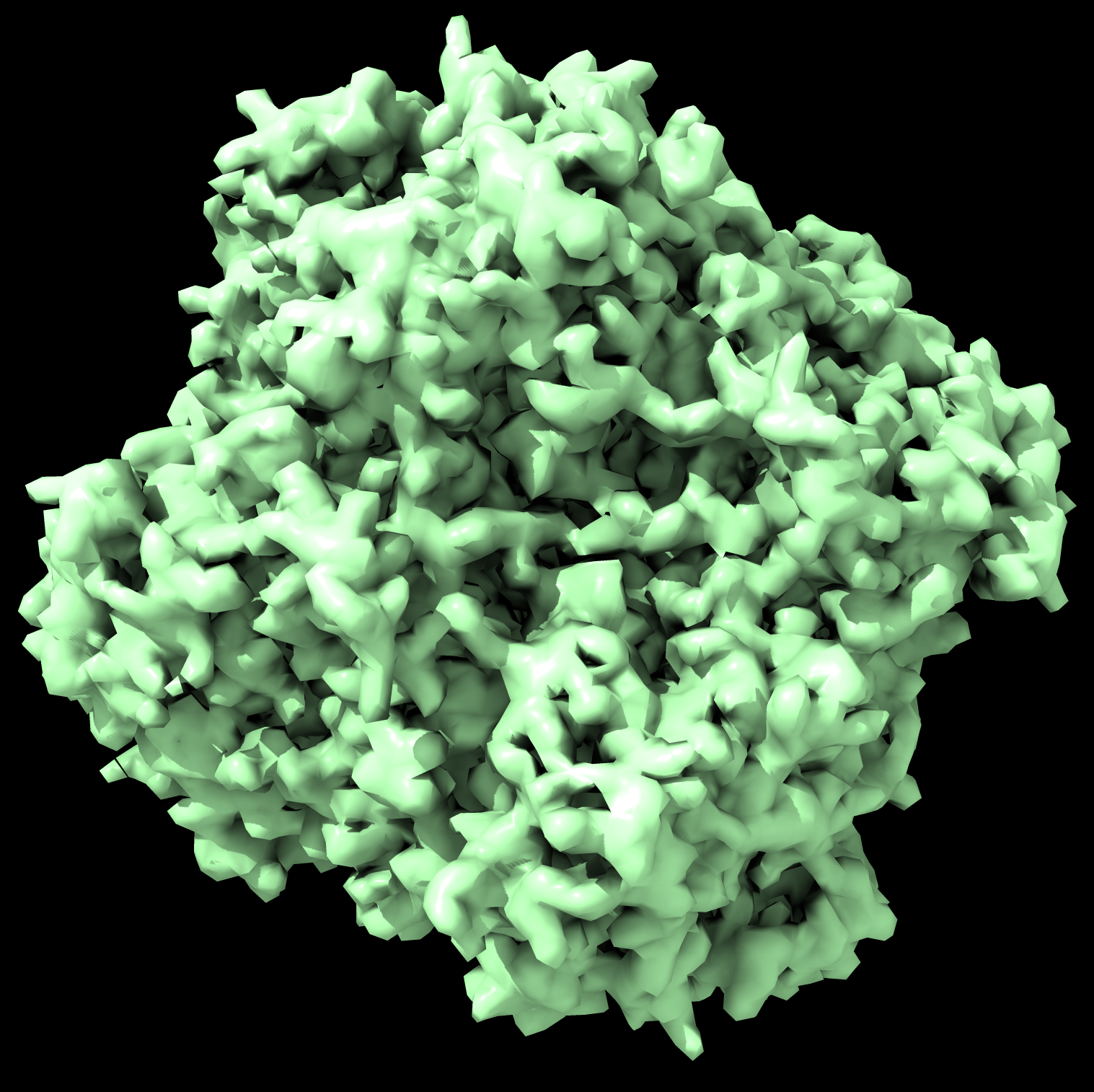}
    \end{minipage}
    \hfill
    \begin{minipage}[b]{\lengthproteinspic}
        \centering
        \includegraphics[width=\linewidth]{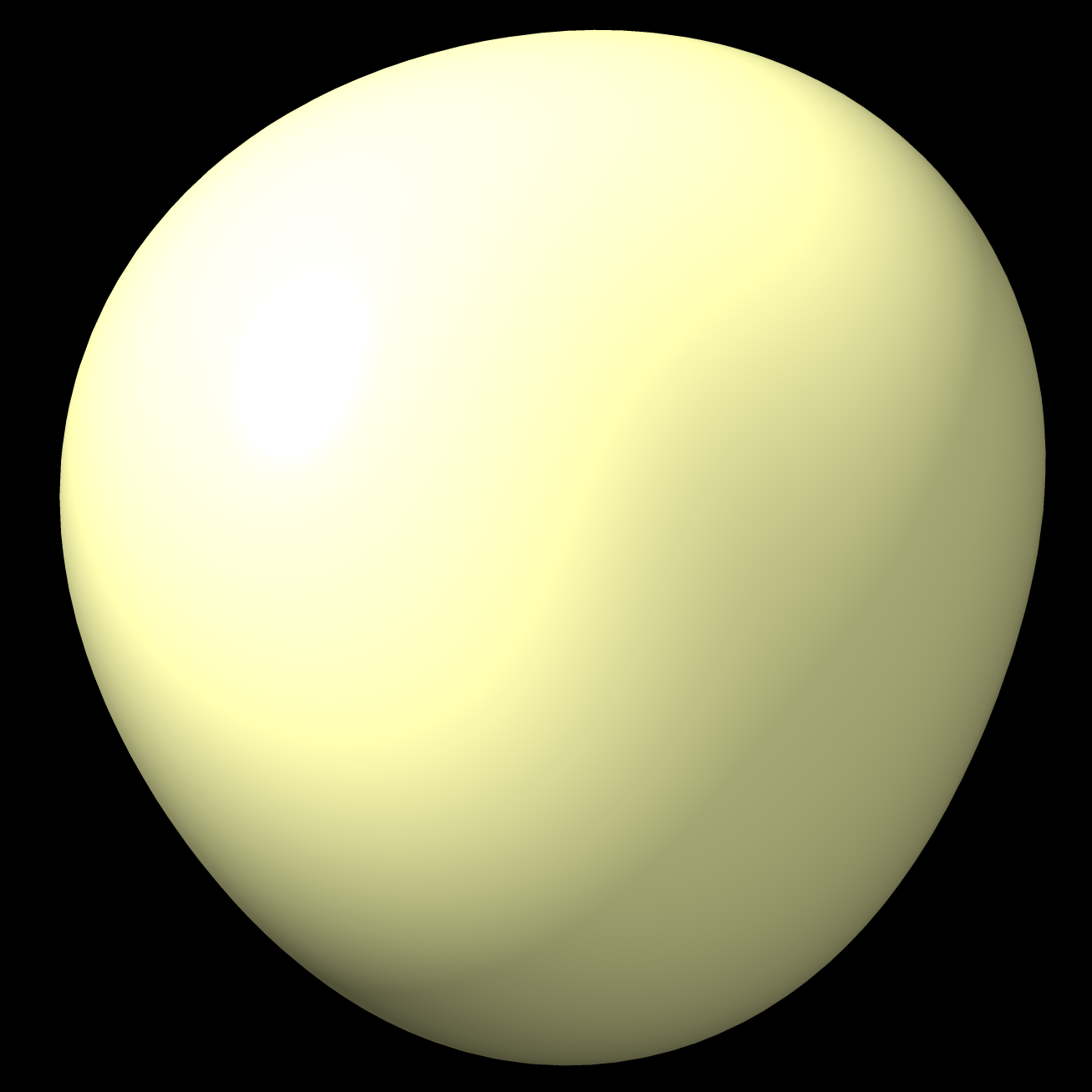}
        (c)
        \vspace{1mm}
        
        \includegraphics[width=\linewidth]{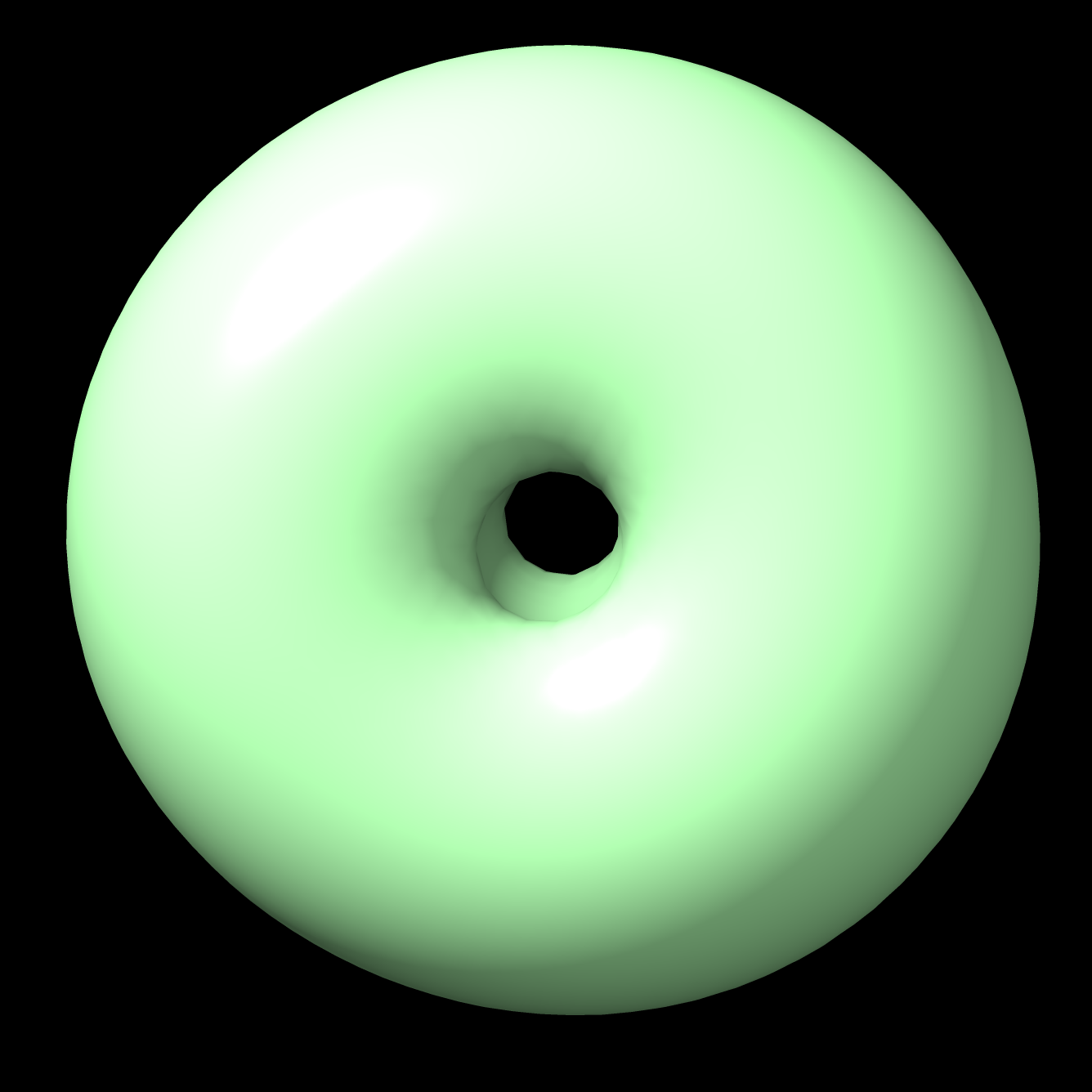}
    \end{minipage}
    \hfill
    \begin{minipage}[b]{\lengthproteinspic}
        \centering
        \includegraphics[width=\linewidth]{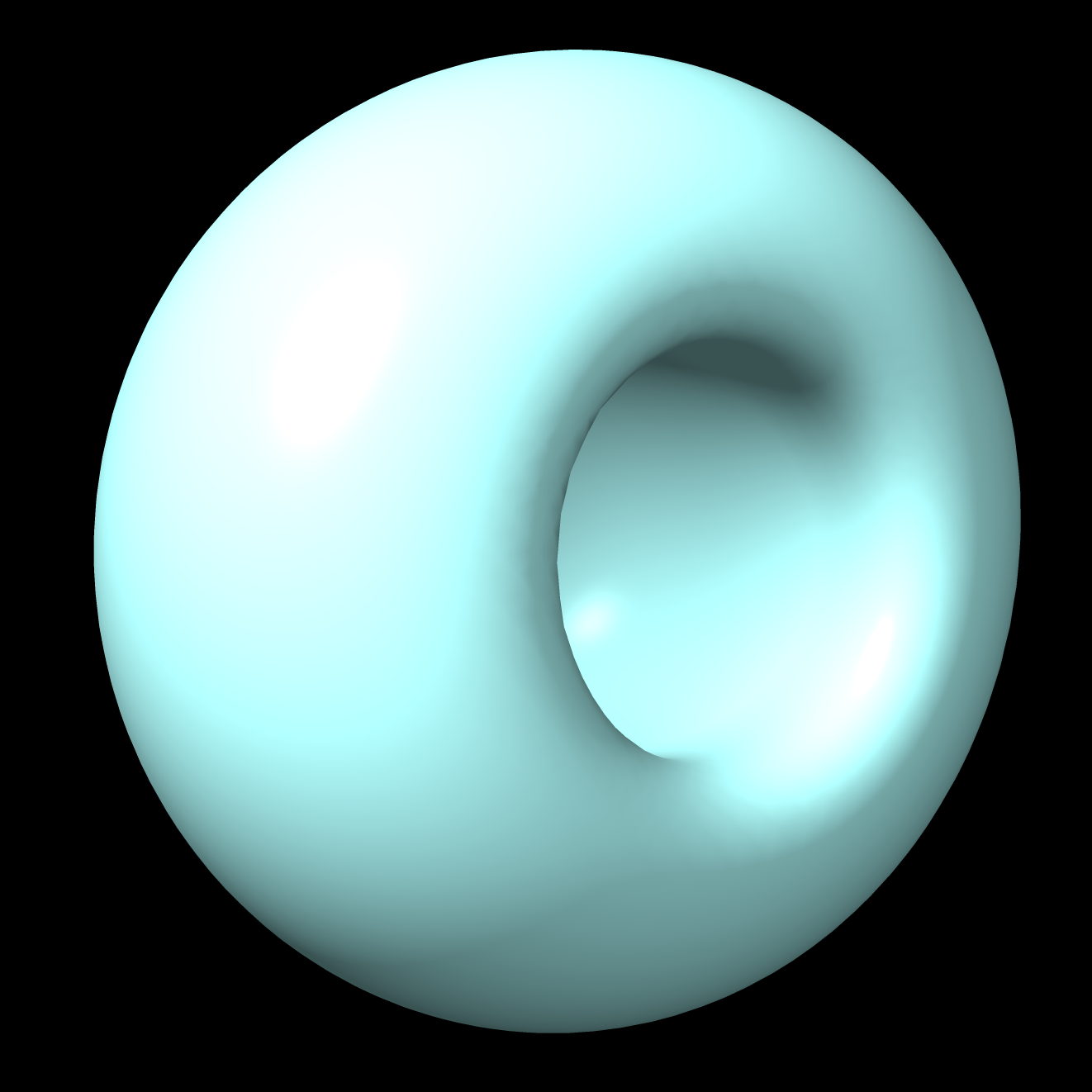}
        (d)
        \vspace{1mm}
        
        \includegraphics[width=\linewidth]{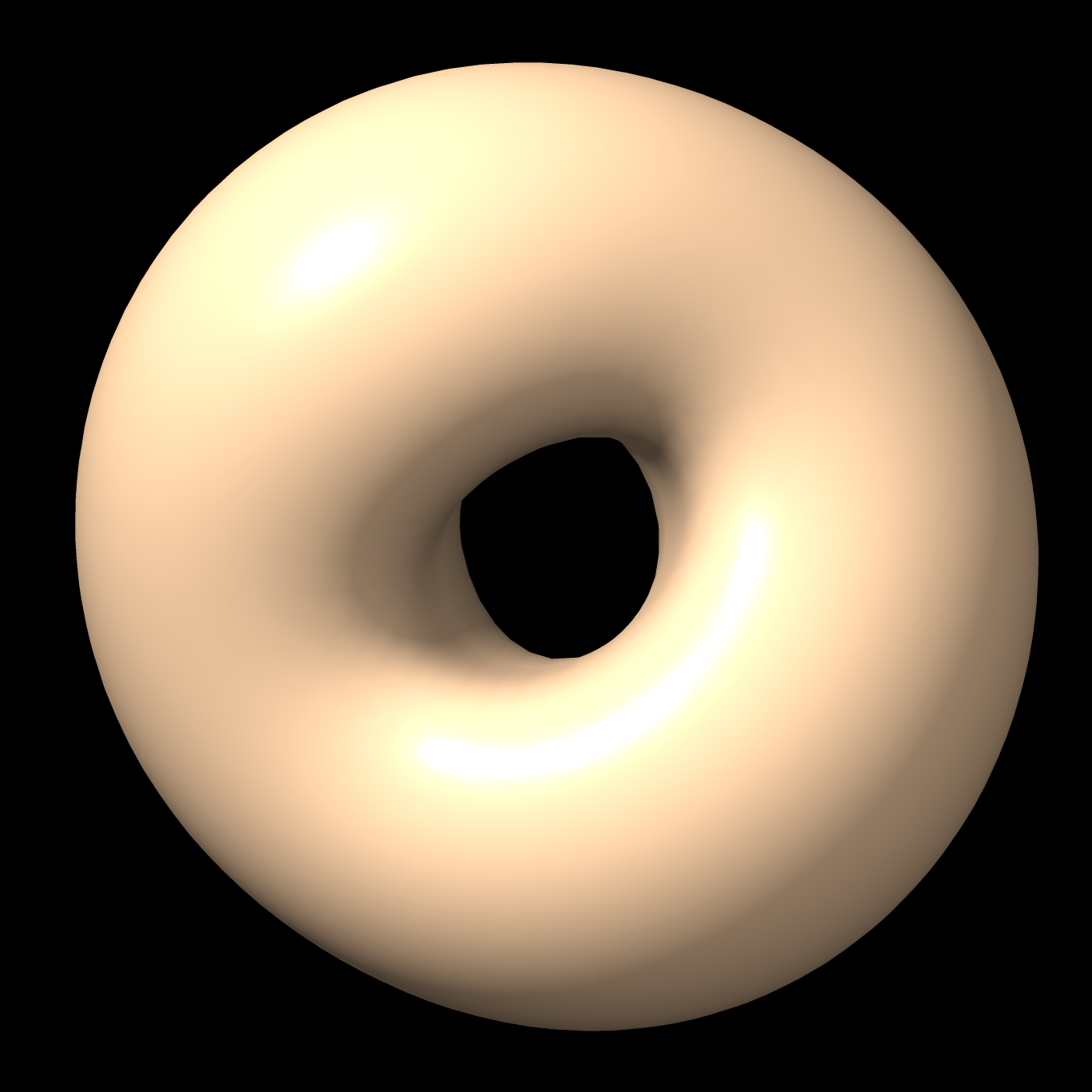}
    \end{minipage}
    \hfill
    \begin{minipage}[b]{\lengthproteinspic}
        \centering
        \includegraphics[width=\linewidth]{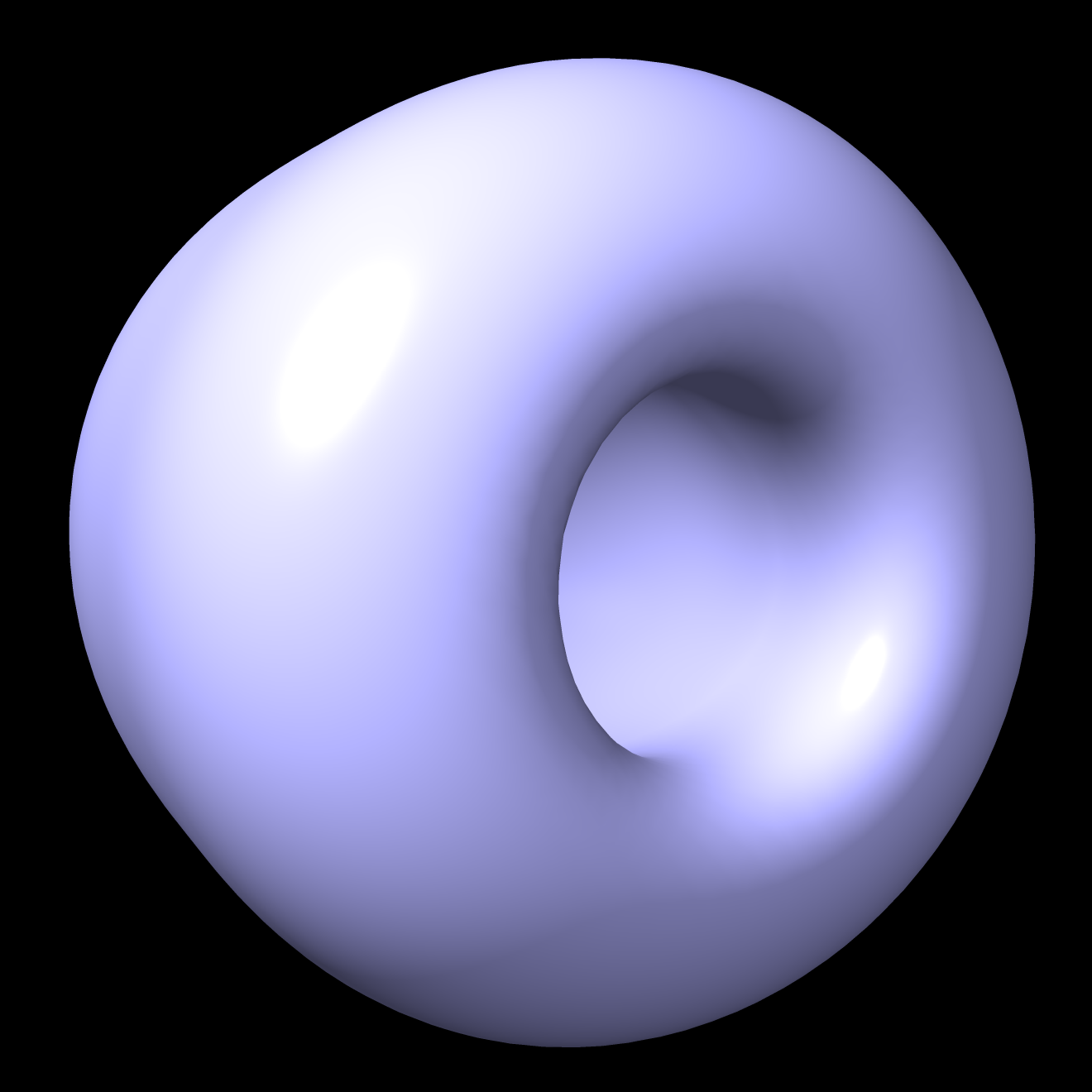}
        (e)
        \vspace{1mm}
        
        \includegraphics[width=\linewidth]{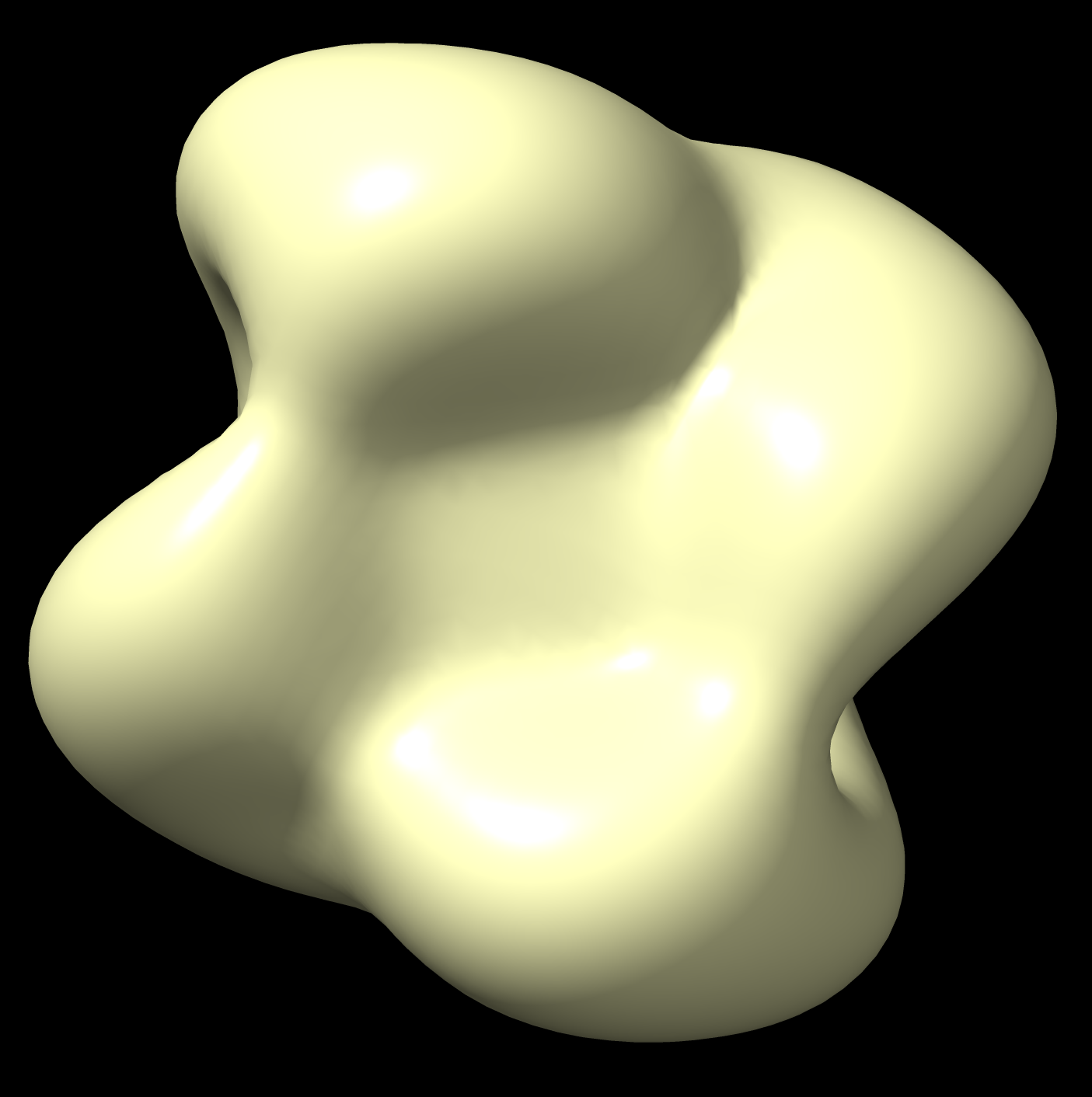}
    \end{minipage}
    \hfill
    \begin{minipage}[b]{\lengthproteinspic}
        \centering
        \includegraphics[width=\linewidth]{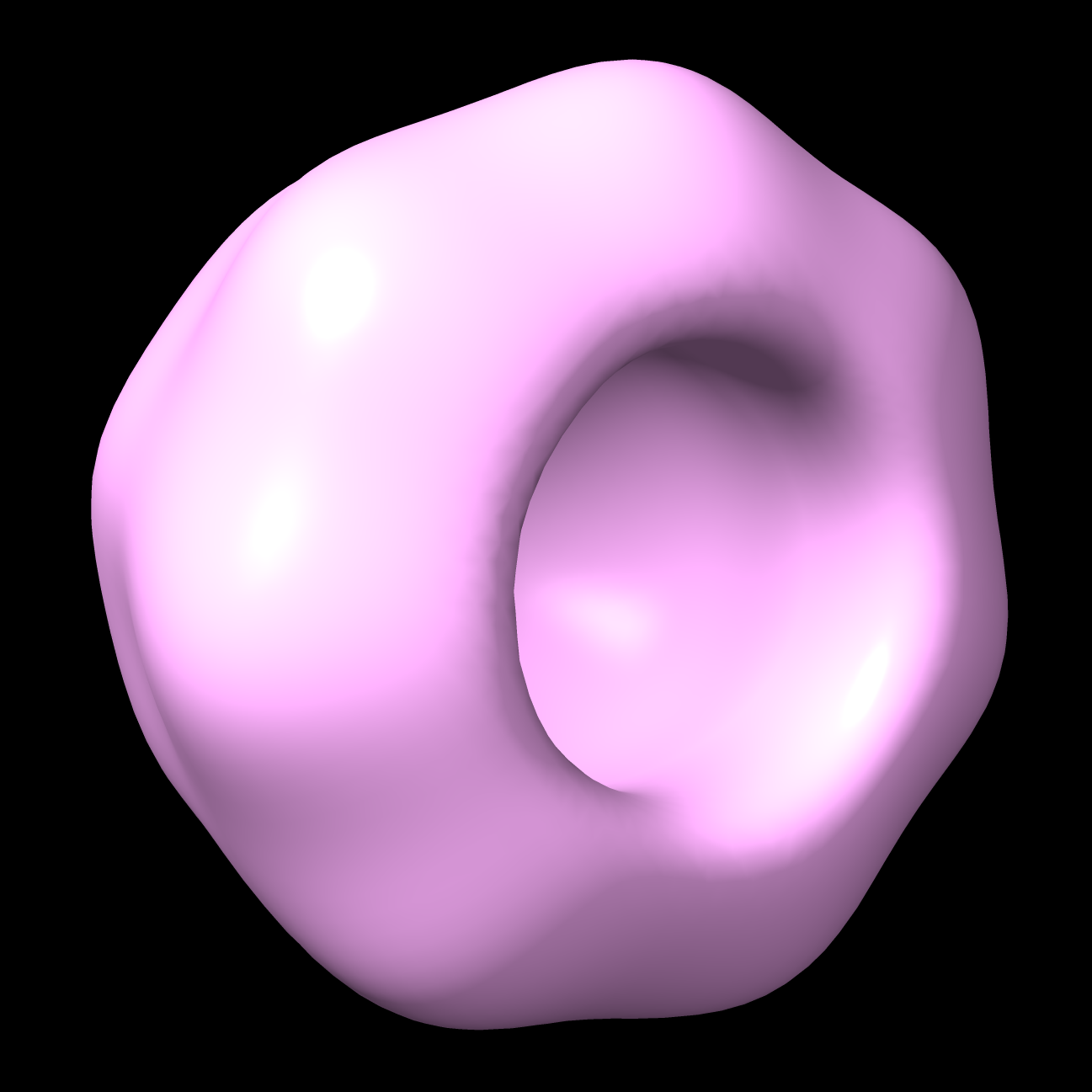}
        (f)
        \vspace{1mm}
        
        \includegraphics[width=\linewidth]{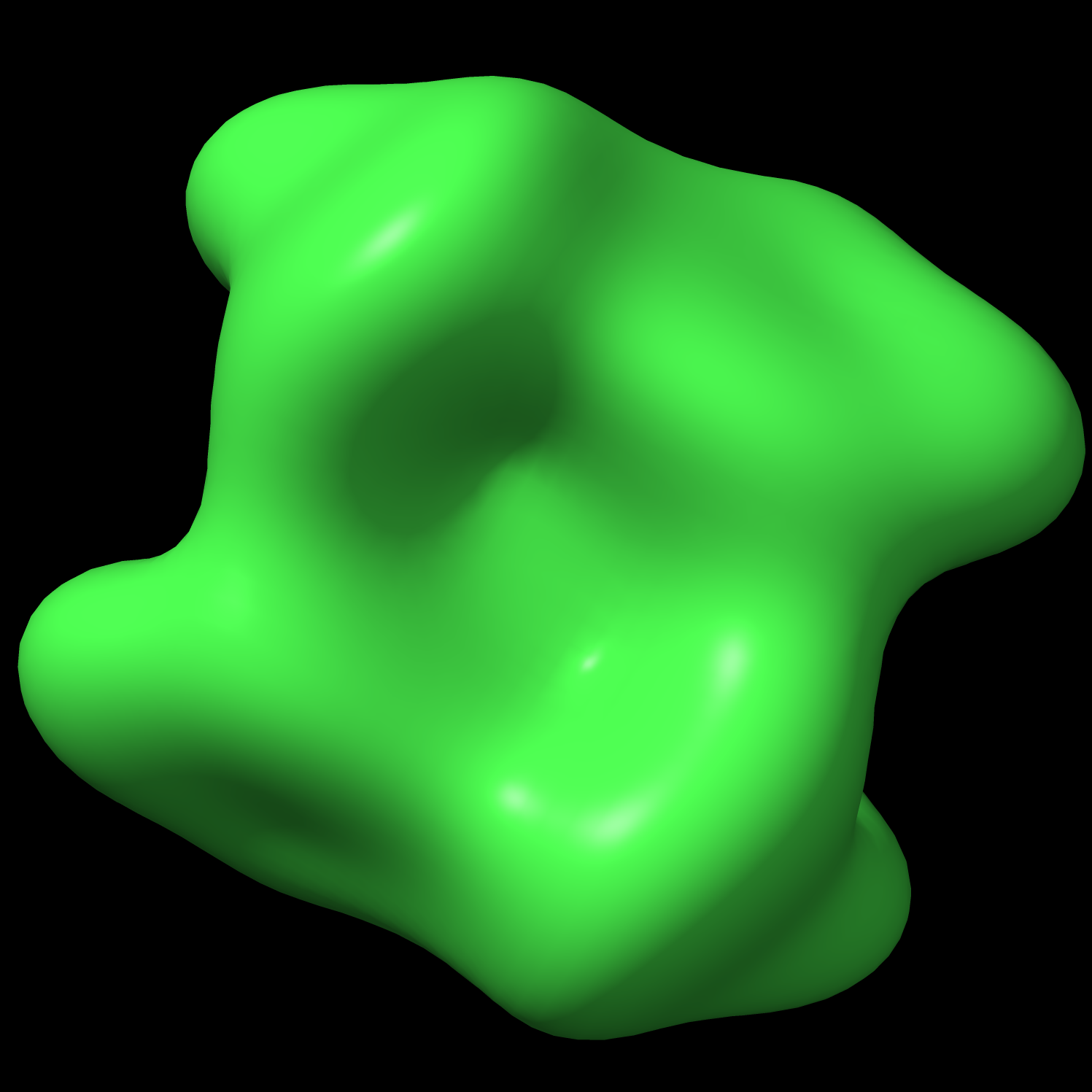}
    \end{minipage}

    \caption{Volume reconstructions with $N=128$.
        In each row:
        (a) reference volume;
        (b) ball harmonics expansion (bandlimit $L=20$);
        (c--f) reconstructions using the top $d$ eigenvectors with $d = 10,20,100,200$, respectively.
        Top row: volume with PDB index \texttt{1avo}; bottom row with \texttt{1dgb}.}
    \label{fig:pacrecon}
\end{figure*}

First we compute a collection of  eigenvolumes for varying parameters for volumes of size  $N = 64$.  See~Figure~\ref{fig:pcaeigenv} for the results.
 Figure~\ref{fig:pacrecon} shows  two molecular structures and their approximation based on the $d$ leading principal volumes.

\begin{figure*}[p]
    \centering

    \begin{minipage}[b]{0.24\linewidth}
        \centering
        \includegraphics[width=\linewidth]{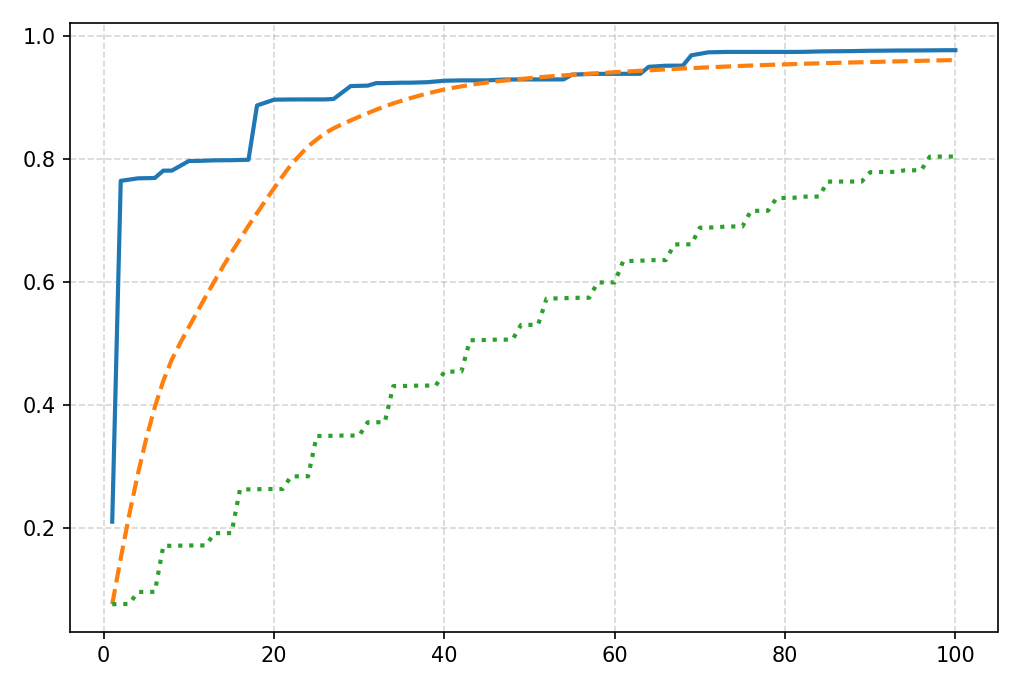}
        (a) $N = 64$, $d = 100$.
    \end{minipage}
    \hfill
    \begin{minipage}[b]{0.24\linewidth}
        \centering
        \includegraphics[width=\linewidth]{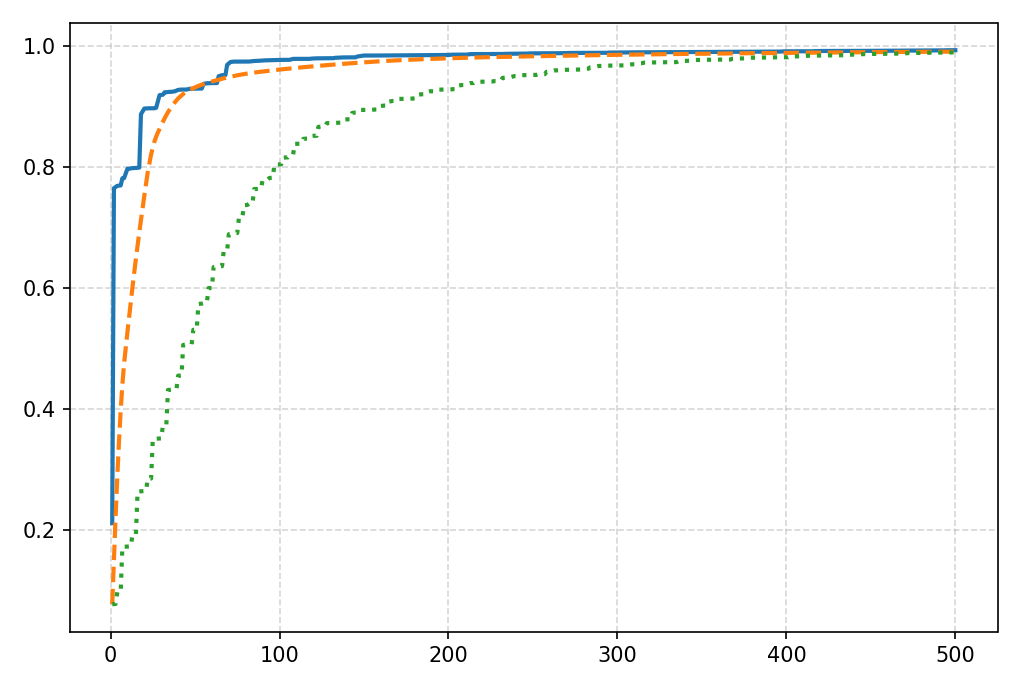}
        (b) $N = 64$, $d = 500$.
    \end{minipage}
    \hfill
    \begin{minipage}[b]{0.24\linewidth}
        \centering
        \includegraphics[width=\linewidth]{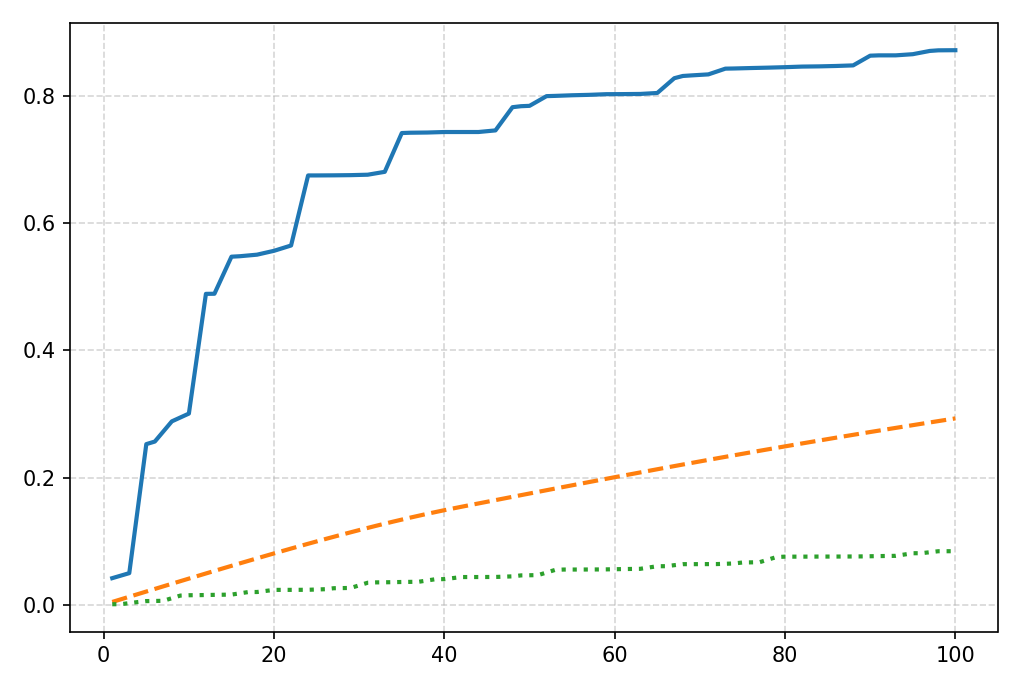}
        (c) $N = 256$, $d = 100$.
    \end{minipage}
    \hfill
    \begin{minipage}[b]{0.24\linewidth}
        \centering
        \includegraphics[width=\linewidth]{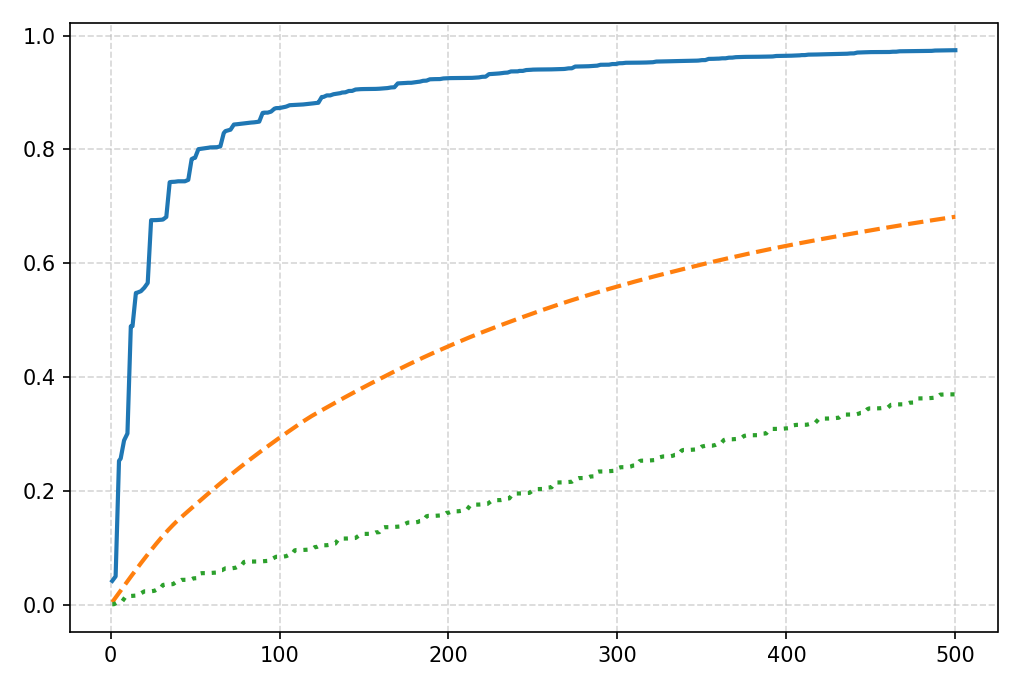}
        (d) $N = 256$, $d = 500$.
    \end{minipage}

    \caption{Comparison of approximations with $\SO(3)$-invariant PCA and the ball harmonics basis (BH) for the sample volume \texttt{1fzf}.
    The plots show $w^V_\phi (k)$ for $k = 1,...,d$ under three choices of the basis $V$: PCA (solid blue), sorted BH (dashed orange), BH sorted by $u_{ls}$ (dotted green).}

    \label{fig:pcavsbes}
\end{figure*}

 \textbf{Comparison Against a Fixed Basis.}
Next, we aim to show that the data-driven PCA approximation outperforms the fixed ball harmonics basis when it is truncated.
We quantify the accuracy of an approximation of a structure $\phi$ by
\begin{equation} \label{intro:energy}
	w_\phi^V (d) \coloneq \frac{ \sum_{j=0}^d |\alpha_j|^2 }{ \sum_{j=0}^D |\alpha_j|^2 } ,
\end{equation}
where $\{ \alpha_j \}_{j=1}^D$ are the expansion coefficients of $\phi$ in an orthonormal basis $V$. 
Figure~\ref{fig:pcavsbes} compares $w_\phi^{V}(d)$ for a representative volume contained in the dataset under three choices of basis~$V$:
the PCA basis~$\mathbf{V}$,
    the ball harmonics  basis with coefficients ordered by decreasing absolute value (this ordering is specific to each volume),
    and the ball harmonics basis with coefficients ordered by increasing $u_{ls}$ (the $s$-th positive root of $j_l$) with near-zero coefficients filtered out.
Two conclusions are evident. First, adaptive PCA offers a clear advantage over a fixed basis.
Second, for volumes sampled on an $N \times N \times N$ grid, the accuracy of the PCA approximation remains nearly unchanged as $N$ increases. In contrast, that of the ball harmonics basis declines sharply, even when the expansion coefficients are reordered by decreasing \nolinebreak magnitude.

\begin{figure*}[p]
    \centering

    \begin{minipage}[b]{0.16\linewidth}
        \centering
        \includegraphics[width=\linewidth]{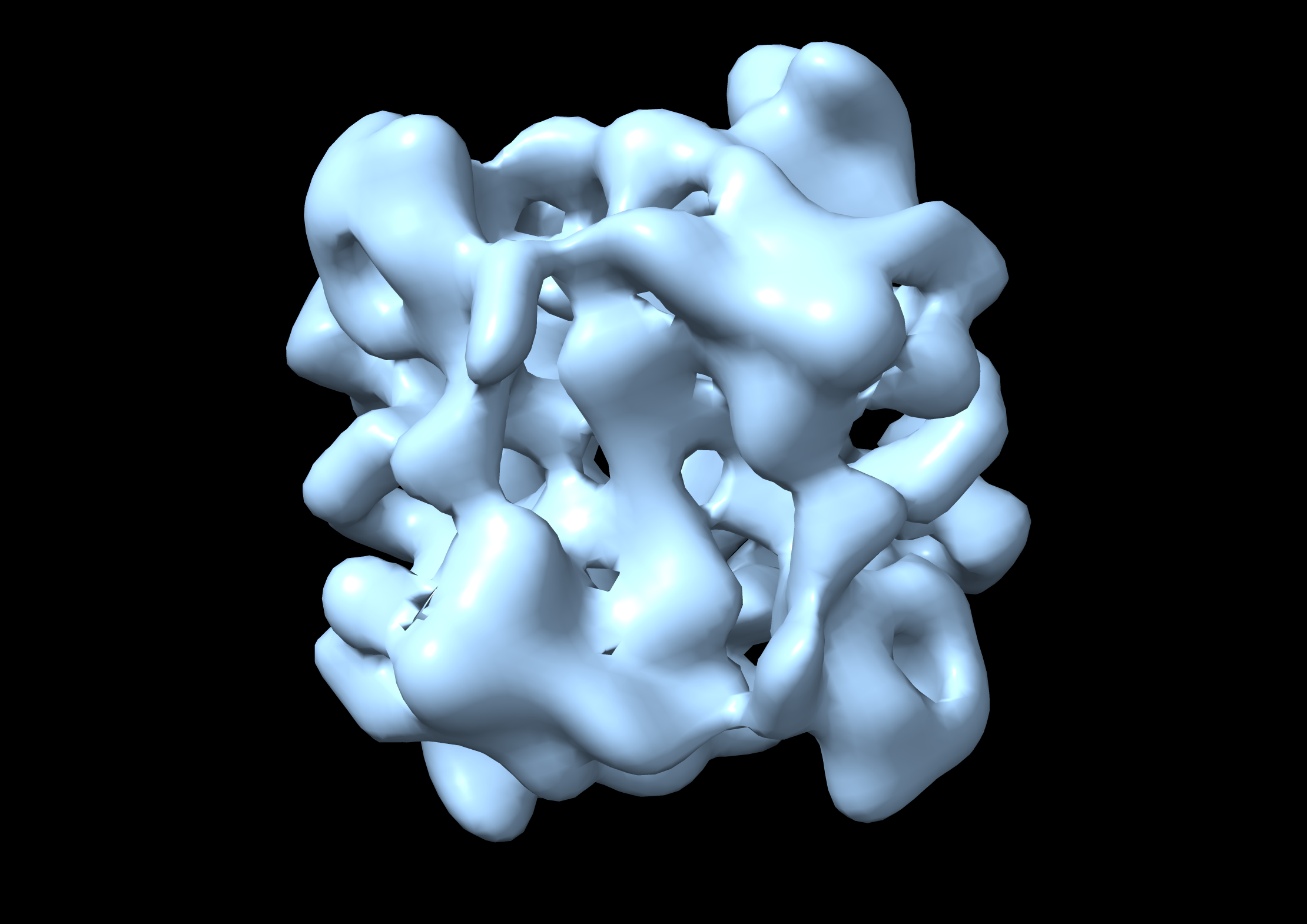}
    \end{minipage}
    \hfill
    \begin{minipage}[b]{0.16\linewidth}
        \centering
        \includegraphics[width=\linewidth]{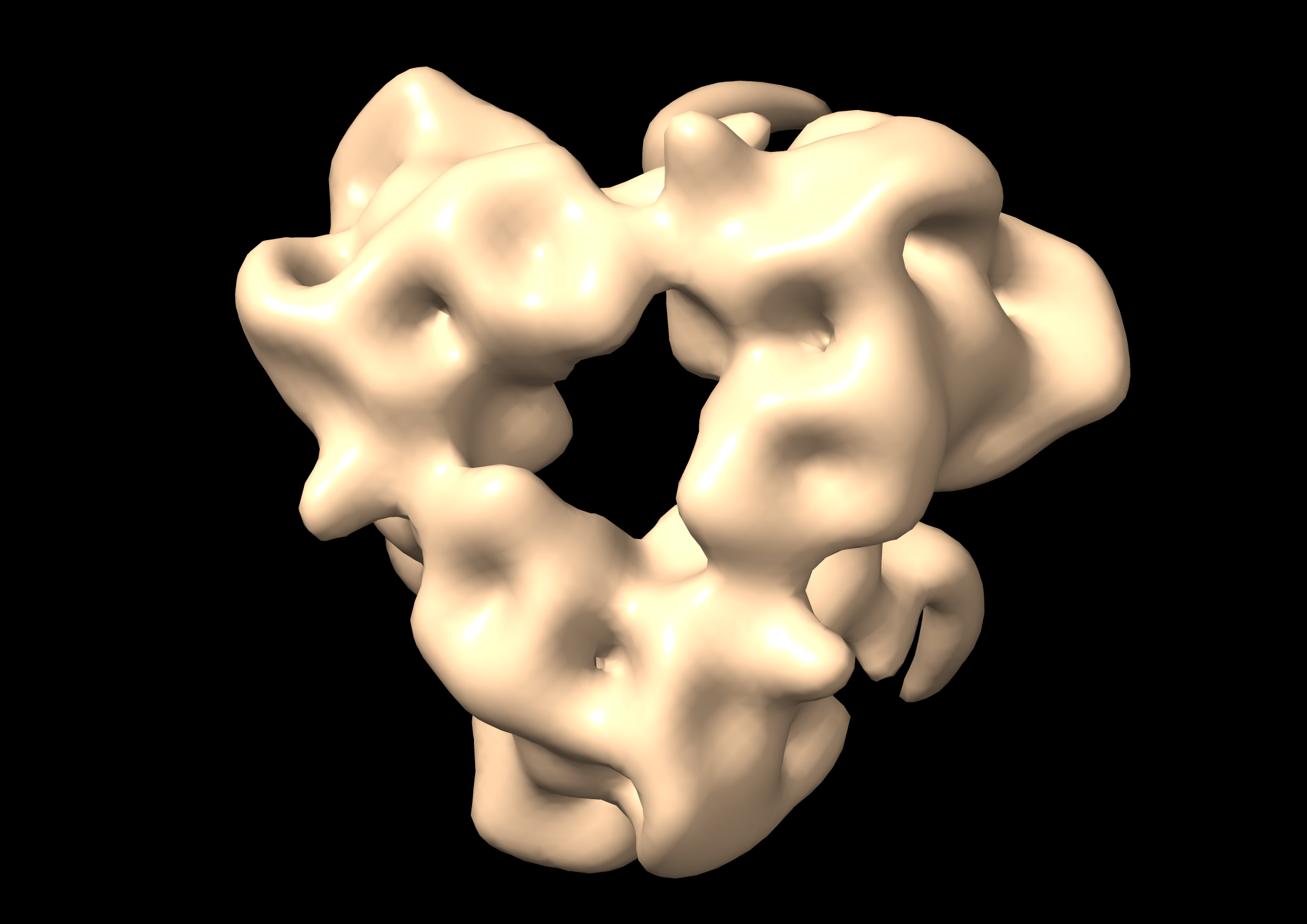}
    \end{minipage}
    \hfill
    \begin{minipage}[b]{0.16\linewidth}
        \centering
        \includegraphics[width=\linewidth]{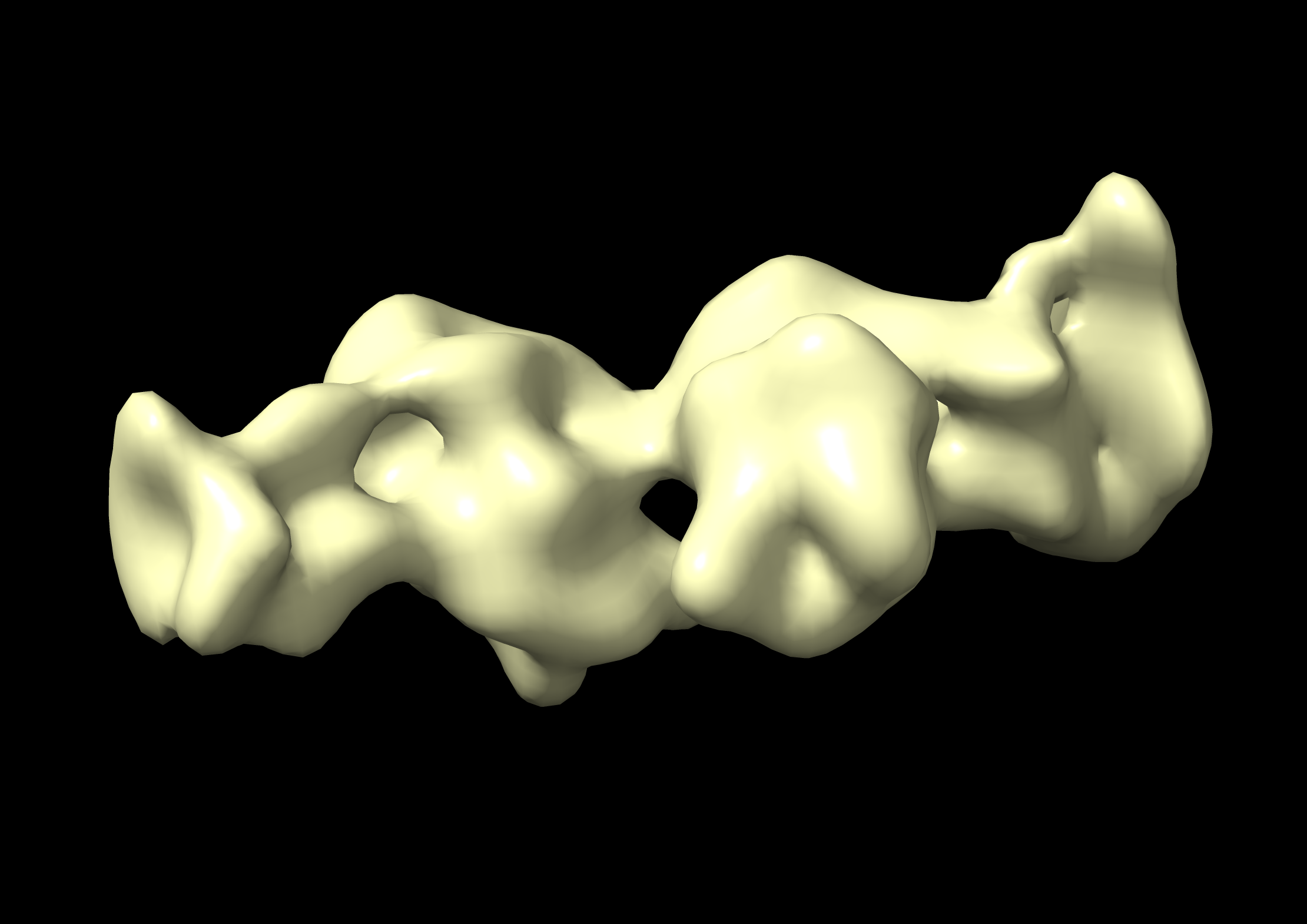}
    \end{minipage}
    \hfill
    \begin{minipage}[b]{0.16\linewidth}
        \centering
        \includegraphics[width=\linewidth]{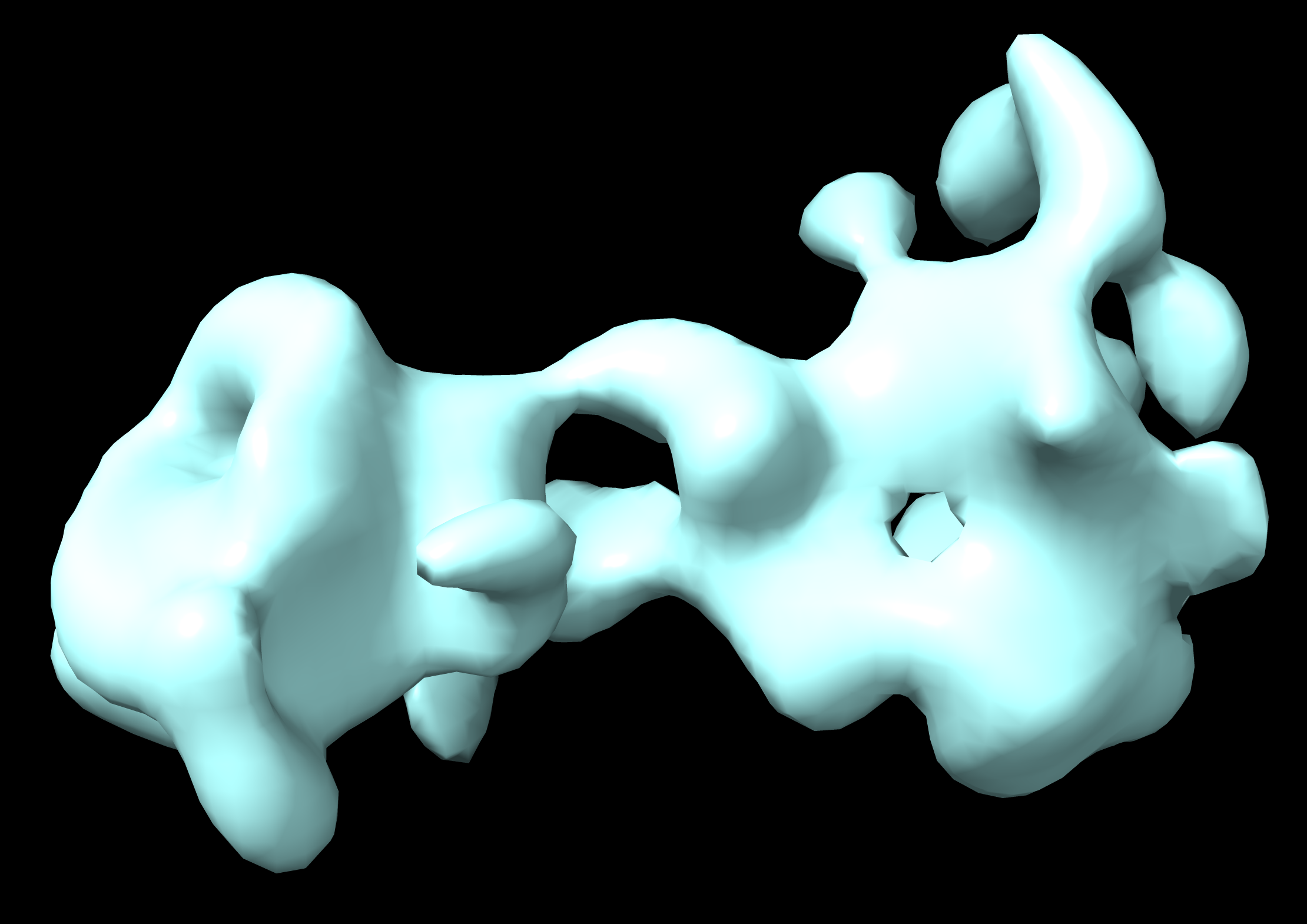}
    \end{minipage}
    \hfill
    \begin{minipage}[b]{0.16\linewidth}
        \centering
        \includegraphics[width=\linewidth]{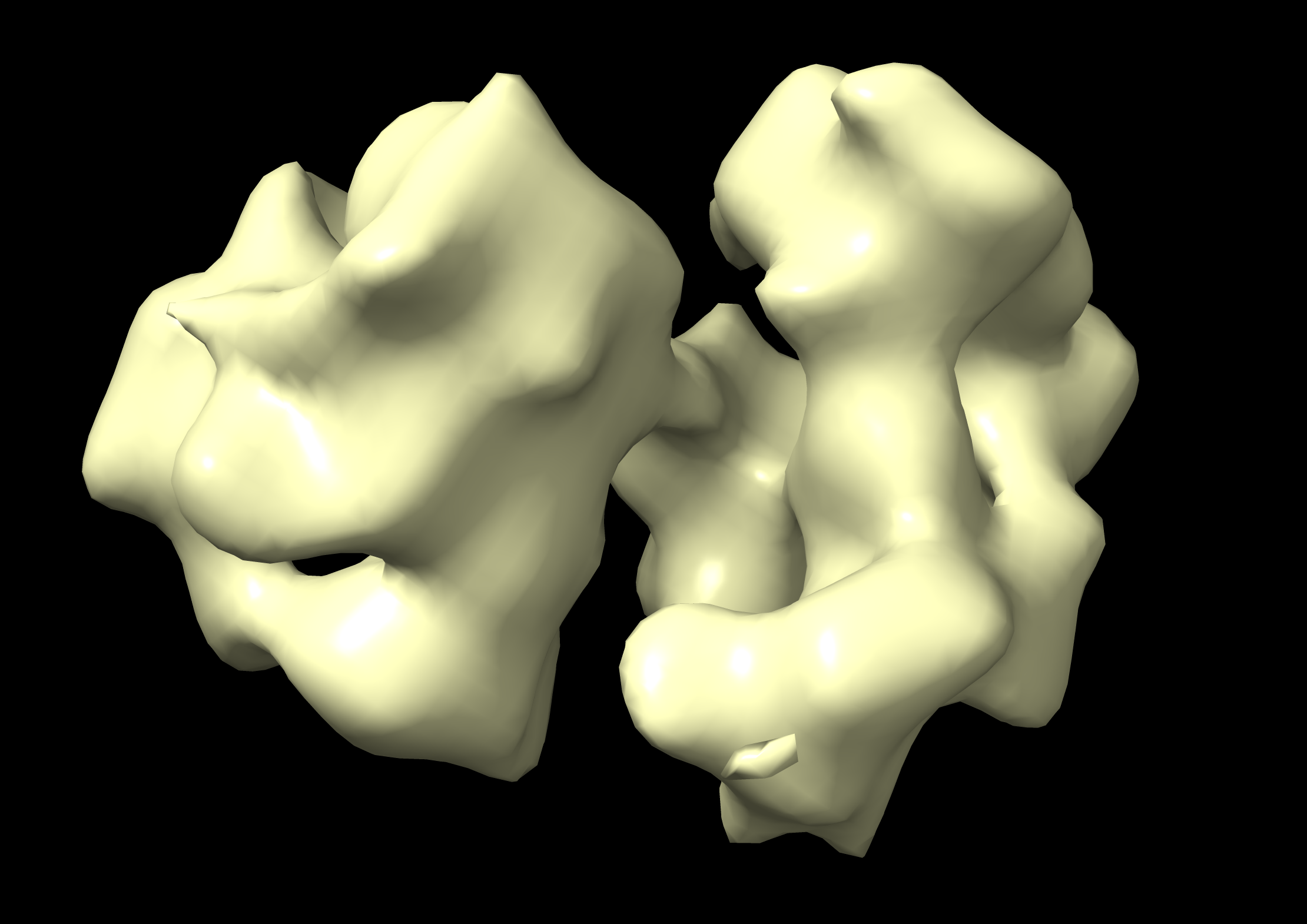}
    \end{minipage}
    \hfill
    \begin{minipage}[b]{0.16\linewidth}
        \centering
        \includegraphics[width=\linewidth]{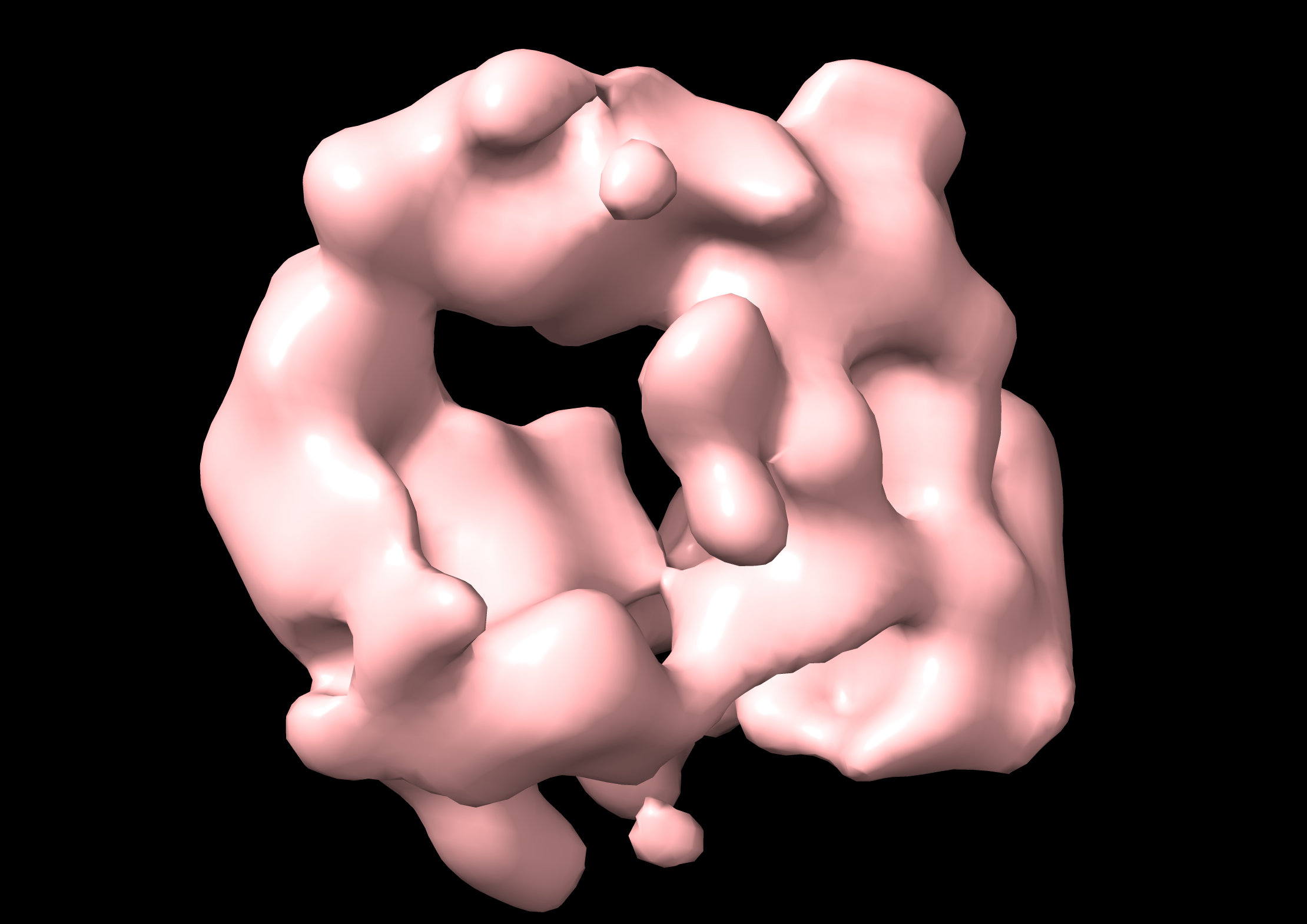}
    \end{minipage}

    \begin{minipage}[b]{0.49\linewidth}
        \centering
        (a) Approximations of actual proteins
    \end{minipage}
    \hfill
    \begin{minipage}[b]{0.49\linewidth}
        \centering
        (b) Synthesized proteins
    \end{minipage}

    \caption{Examples of synthesis of random proteins using $d = 200$ eigenvectors with resolution $N = 64$ and bandlimit $L = 20$.
    The PDB indexes of the real proteins are \texttt{1dgb}, \texttt{1cb5}, \texttt{1fzf}, respectively.} 
    \label{fig:fakeprot}
\end{figure*}

\textbf{Synthesizing Proteins.}
PCA can be thought of as giving a generative model. 
Generative models are useful for data augmentation in modern machine learning because they can create new, realistic examples that mimic the true data distribution. 
This may be especially valuable for structural biology, where data is expensive to collect. 
With PCA, approximations of the molecular structures lie in a linear space $\mathbf{L} \cong \mathbb{R}^d \subseteq \mathbb{R}^D$.
We suggest synthesizing other elements of~$\mathbf{L}$ (that is, “fake proteins”) in the following way.
Let $\Phi$ be the dataset of proteins and let $\{\psi_j\}_{j=1}^d$ be its low-rank basis obtained via the $\SO(3)$-invariant PCA procedure.
For each ${\phi^{(i)} \in \Phi}$, we consider its expansion $\phi^{(i)} = \sum_{j=1}^d \alpha_{j}^{(i)} \psi_j$.
The mean $\mu_j$ and the variance $\sigma_j$ are estimated so that the samples $\{ \alpha_{j}^{(i)} \}_{\phi^{(i)} \in \Phi}$ are modeled as $\mathcal{N} (\mu_j, \sigma_j^2)$.
Now, a random element of $\mathbf{L}$ can be sampled as $\sum_{j=1}^d \beta_j \psi_j$,
    where $ \beta_j  \sim \mathcal{N} ( \mu_j, \sigma_j^2 )$ is a random sequence of coefficients.
We present a few examples of synthesized proteins in Figure~\ref{fig:fakeprot}, compared to some real ones.

\vspace*{-1mm}
\section{Pathways to Applications in cryo-EM} \label{sec:applications}

This paper derived an efficient  method for performing $\SO(3)$-invariant PCA on 3D data samples with random orientations. 
To conclude, we outline potential applications of the framework, with emphasis on 3D reconstruction in cryo-EM. 

The main computational workhorse in cryo-EM is the expectation–maximization (EM) algorithm~\cite{scheres2012relion}. 
Although it is 
successful in recovering a wide range of molecular structures, there is strong evidence that its performance deteriorates as the noise level increases~\cite{balanov2025expectation}. 
One possible 
remedy is to apply EM directly on a subspace where the molecular structure is expected to lie, a strategy known as subspace EM~\cite{dvornek2015subspaceem}.
Using the computational framework developed in this work, this subspace can be learned from molecular databases and incorporated into existing EM pipelines.   
We note that implementing this will present challenges, e.g., due to CTF effects~\cite{micksingfastpca}.

As an alternative to EM, the method of moments (MoM) has been proposed 
for cryo-EM, 
particularly for scenarios with extremely high noise. 
This approach is auspicious for small molecular structures, which typically induce 
high noise levels and 
remain beyond the reach of current cryo-EM technology~\cite{bendory2023toward}. Recent work has shown that  
MoM can be improved when the structure is constrained to a subspace~\cite{hoskins2024subspace}.
The subspace could be learned using the framework developed in this paper, after which moments would be expressed in terms of expansion coefficients in the resulting basis.

\section*{Acknowledgments}
T.B. and N.S. are supported in part by the NSF-BSF under Grant 2024791. T.B. is also supported in part by BSF under Grant 2020159, in part by ISF under Grant 1924/21, and in part by a grant from The Center for AI and Data Science at Tel Aviv University. N.S. is also partially supported by the BSF award 2024266 and the DFG award 514588180.
J.K. is supported in part by NSF DMS 2309782, NSF DMS 2436499, NSF CISE-IIS 2312746, DE SC0025312, and the Sloan Foundation.
A.S. is supported in part by AFOSR FA9550-23-1-0249, the Simons Foundation Math+X Investigator Award, NSF DMS 2510039, and NIH/NIGMS R01GM13678001.

\bibliographystyle{IEEEbib}

\end{document}